\documentclass[aps,eqsecnum]{revtex4}
\usepackage{epsfig,rotating}
\newcommand{\be}{\begin{equation}}
\newcommand{\ee}{\end{equation}}
\newcommand{\bea}{\begin{eqnarray}}
\newcommand{\eea}{\end{eqnarray}}

\begin{document} 
\title{Tsunami Inflation: Selfconsistent Quantum
Dynamics\footnote{Lectures at the 
8th. Erice Chalonge School, p. 45-80 in the Proceedings, edited by H. J. de
Vega, I. M. Khalatnikov and N. G. Sanchez, vol 40, series II, Kluwer 2001.}}
\author{H. J. de Vega}
\affiliation{LPTHE, Universit\'e
Pierre et Marie Curie (Paris VI) et Denis Diderot  
(Paris VII), Tour 16, 1er. \'etage, 4, Place Jussieu, 75252 Paris
cedex 05, France}
\begin{abstract}  
The physics during the inflationary stage of the universe is of
quantum nature involving extremely high energy densities. Moreover, it
is out of equilibrium on a fastly expanding dynamical geometry.
We complement here the 1999 Chalonge Lectures on out of
equilibrium fields in self-consistent inflationary dynamics [astro-ph/0006446]
investigating inflation driven by the evolution of highly
excited {\bf quantum states}. These states are characterized by a
non-perturbatively large 
number of quanta in a band of momenta and with zero or nonzero
expectation value of the inflaton scalar field. They represent the situation in
which initially a non-perturbatively large energy density is localized 
in a band of high energy quantum modes and are  coined tsunami-waves.
The self-consistent evolution of this quantum state and the scale
factor is studied analytically and numerically. It is shown that the
time evolution of these  quantum states lead to two consecutive stages 
of inflation under conditions that are the quantum analogue of
slow-roll. The evolution of the scale factor during the first stage
has new features that are characteristic of the quantum state.  During
this initial stage the quantum fluctuations in the highly excited band 
build up an effective homogeneous condensate with a non-perturbatively
large amplitude as a consequence of the large number of quanta. 
The second stage of inflation is similar to the usual classical
chaotic scenario 
but driven by this effective condensate. The excited quantum 
modes are already superhorizon in the first stage and do not affect the
power spectrum of scalar perturbations.  Thus, this 
tsunami quantum state provides a field theoretical justification
for chaotic scenarios driven by a classical homogeneous 
scalar field of large amplitude. 
\end{abstract}
\maketitle
\tableofcontents

\section{Introduction}

A wealth of observational evidence from the temperature anisotropies
in the cosmic microwave background strongly points towards inflation as the
mechanism to produce the primordial density 
perturbations\cite{turner,cmb}. Thus, inflationary cosmology 
emerges as the basic theoretical framework to explain not only the
long-standing shortcomings  of standard big bang 
cosmology but also to provide a testable paradigm for structure
formation\cite{revius}-\cite{infl}. The recent explosion in
the quantity and quality of data on temperature anisotropies 
elevates inflation to the realm of an experimentally testable scenario
that leads to robust predictions that 
withstand detailed scrutiny\cite{turner,cmb}. 

However at the level of implementation of an inflationary proposal,
the situation is much less satisfactory. There 
are very many different models for inflation motivated by particle
physics and most if not all of them invoke one or several scalar 
fields, the inflaton(s), whose dynamical evolution in a scalar
potential leads to an inflationary epoch\cite{revius}-\cite{infl}. 
The inflaton field is a scalar field that provides an effective
description for the fields in the grand unified theories. Furthermore
there is the tantalizing prospect of learning some aspects of the
inflationary potential (at least the part of the potential associated with  
the last few e-folds) through the temperature anisotropies of the
cosmic microwave background\cite{reconstruction}.  

Most treatments of inflation study the evolution of the inflaton via
the {\em classical} equations of motion for the chosen scalar
potential instead of making a quantum field treatment of the
dynamics. That is, the effect of quantum fluctuations is neglected in the 
dynamics of the inflaton. Furthermore, since inflation redshifts  
inhomogeneities very fast, the classical evolution is studied in terms
of a {\em homogeneous classical scalar} field. The quantum 
field theory interpretation is that this classical, homogeneous field
configuration is the expectation value of a quantum field operator 
in a translational invariant quantum state. While the evolution of
this coherent field configuration (the expectation value or  
order parameter) is studied via classical equations of motion,
quantum fluctuations of the scalar field around this expectation
value are treated in a linear approximation for the high wavenumbers
that lead to the  seeds for scalar density perturbations of the
metric\cite{revius}-\cite{infl}.  The large amplitude modes that
dominate the energy  of the universe during inflation are mimic by the
classical homogeneous field. Moreover, the slow roll approximation is
used in all cases. 

A fairly broad
catalog of inflationary models based on scalar field dynamics labels
these either as `small field' or  
`large field'\cite{reconstruction}. In the `small field' category  the
scalar field begins its evolution with an initial value very near the
origin of the scalar potential and rolls 
down towards larger values, an example is new
inflation\cite{revius,coles,linde}. In the `large field' category, the scalar field
begins very high up in the potential hill and rolls down 
towards smaller values, an example is chaotic
inflation\cite{revius,coles,linde}.  

It is only recently that the quantum dynamics of the
scalar fields in the coupled evolution of matter and geometry has
been studied self-consistently\cite{noscos,asam}. This work is
associated with the dynamics of non-equilibrium phase transitions in models 
that fall, broadly, in the `small field' category. 

This subject is
reviewed in the 1999 Chalonge Lectures on out of equilibrium fields in
self-consistent inflationary dynamics. The conclusion of these
works is that a  treatment of the quantum fluctuations that couple
self-consistently to the dynamics of the metric 
provides a solid quantum field theoretical framework that justifies
microscopically the picture based on classical inflation.  
At the same time these studies provide a deeper understanding of the
quantum as well as classical aspects of inflation and inflationary 
perturbations. They clearly reveal the classicalization of initial
quantum fluctuations\cite{noscos,asam}, and furnish a microscopic
explanation (and derivation) of the effective, homogeneous classical
inflaton\cite{asam}.   

Very recently, a quantum dynamical  treatment of
models whose classical counterpart are large field models is proposed
in ref.\cite{tsuinf}. The {\em classical}
description in these models begins with a homogeneous inflaton scalar with very
large amplitude $\phi \sim M_{Pl}$\cite{revius,coles,linde,infl}, i.e,
very high up  in the scalar potential well. 

The initial quantum state may be mixed (described by a density matrix
in Fock space) or a pure state (described by a vector in Fock space).
The initial state is characterized by the expectation value of the
inflaton (order parameter) and the spectrum of excitations (initial
particle distribution). The order parameter corresponds to the
classical inflaton field in the classical limit whereas the initial
particle distribution describes the spectrum of excitations in the
initial state. 

It is not obvious whether an initial particle distribution
can give rise to inflation. As shownin ref.\cite{tsuinf}, there exists
classes of initial particle distributions leading to efficient inflation.
In particular, initial states with zero order parameter and a band of
excited modes can lead to an inflationary epoch. This extends the set
of possible initial states that leads to chaotic inflation making it a
more natural description of the early universe. 

In the customary treatment of chaotic inflation (classical chaotic inflation 
\cite{revius,infl}) all the energy is contained in the classical (space
independent) field, so the field modes do not contribute neither to the 
energy nor to the background dynamics. 
This is clearly a very special choice of initial conditions. 
We shall call this scenario classical chaotic inflation. 

We review in these lectures the dynamics that results from the
evolution of a  {\bf quantum state} which drives the 
dynamics of the scale factor through  the expectation value of the
energy momentum tensor\cite{tsuinf}. The main idea behind this 
approach is akin to the experimental situation in ultrarelativistic
heavy ion collisions, wherein an initial highly excited state  
(heavy ions with very large energy) lead to the formation of a plasma
that expands and cools\cite{harris}. Recently  this situation has 
been modeled  by considering the evolution of highly excited
quantum states  coined tsunami-waves in\cite{tsurob,tsu1,tsu2}.

{\em The goals of the Tsunami Inflation:}

The ideas and concepts in refs.\cite{tsurob,tsu1,tsu2} were adapted in
ref.\cite{tsuinf} to study the self-consistent 
dynamics of the metric and the evolution of a highly excited quantum
state with the goal of providing a quantum description of large field
inflationary models {\em without assuming an expectation 
value for the scalar field}. Novel quantum states were considered, which
are the cosmological counterpart of the  tsunami-waves introduced
in\cite{tsurob,tsu1,tsu2} with the following properties\cite{tsuinf}:  

{\em i)} a quantum state with non-perturbatively large number of
quanta in a momentum band, i.e, a large number of high energy excitations. 
We first solve explicitly the case of a narrow band and then we consider 
more general distributions of quanta. 

{\em ii)} vanishing expectation value of the scalar field.  

The rationale behind considering these quantum states is that they
provide a natural description of a situation in which a state 
of large energy density composed mainly of energetic particles
evolves in time. The main idea is to model the 
situation conjectured to drive inflationary cosmology:  that is an
initial state with a large energy density and pressure which in turn
couples to the metric leading to inflation. Efficient inflation should
follow independently of the details of the initial state. In our case,
this means that efficient inflation is to be obtained for different
shapes of the initial particle distribution. That is, we want to avoid
a {\bf fine tuning} of the initial conditions. 

[In the usual studies (classical chaotic inflation) all of the initial
energy is in the classical zero mode while the quantum 
fluctuations are taken to be in the ground state]. 

After inflation, energy transfer from the heavier to lighter particles
results in reheating and particle production that eventually excites
the light sector and leads to a radiation or matter dominated phase.  

The conditions under which a Tsunami quantum state leads to
inflationary dynamics is established and  the self-consistent 
evolution of this quantum state and the space-time metric is studied
in detail\cite{tsuinf}. 

We emphasize that we are {\bf not} proposing here yet a new model of
inflation. Instead we focus on  inflation driven by the evolution of a
{\bf quantum} state, within the framework of familiar models based  
on scalar fields with typical quartic potentials. This is  in contrast
with the usual approach in which the  dynamics 
is driven by  the evolution of a homogeneous {\bf classical} field of
large amplitude.  

{\em Brief summary:} We find that inflation occurs under fairly
general conditions that are the {\em quantum } equivalent of
slow-roll. 
There are {\em two} consecutive but distinct inflationary stages: the
first one is completely determined by the quantum features of 
the state. Even when the expectation value of the scalar field {\em
vanishes at all times } in this quantum state, the dynamics of the
first stage gives rise to the emergence of an {\em effective classical
homogeneous condensate}. The amplitude of the effective condensate is 
non-perturbatively large as a consequence of the non-perturbatively
large number of quanta in the band of excited wavevectors. The second
stage is similar to the familiar classical chaotic scenario, and can be
interpreted as being driven by the dynamics of the effective
homogeneous condensate\cite{tsuinf}. The band of excited quantum modes, if not
superhorizon initially they cross the horizon during the first stage
of inflation, hence they do not modify the power spectrum of scalar
density perturbations on wavelengths that are of cosmological
relevance today. Actually, in the explicit examples worked out in
ref.\cite{tsuinf}, the excited modes are initially superhorizon due to
the generalized slow-roll condition. Therefore, in a very 
well defined manner, tsunami quantum states provide a quantum
field theoretical justification, a microscopic basis, for chaotic
inflation, explaining the classical dynamics of the homogeneous 
scalar field.   

In section II we introduce the quantum state, obtain the renormalized
equations of motion for the self-consistent evolution of 
the quantum state and the scale factor. In section III we  provide
detailed  analytic and numerical studies of the evolution and 
highlight the different inflationary stages. In section IV we
discuss generalized scenarios. The summary of results is presented 
in the conclusions. An appendix is devoted to technical details and
the equations of motion for mixed states.   

\section{Initial state and equations of motion}

As emphasized in the introduction, while most works on inflation treat the
dynamics of the inflaton field at the classical level, we use a 
quantum description of the inflaton. 

We focus on the possibility of inflation through the dynamical {\em
quantum} evolution of  a highly excited initial state with large energy 
density. Consistently with inflation at a scale well below the Planck
energy, we treat the inflaton field describing the matter as a 
quantum field whereas gravity is treated semiclassically. 

The dynamics of the classical space-time metric is determined by the
Einstein equations  with a source term given by the expectation value
of the energy momentum tensor of the quantum inflaton field. The
quantum field evolution is calculated in the resulting metric. 

Hence  we solve  {\em self-consistently} the coupled evolution
equations for the classical metric and the quantum inflaton field.

We assume that the universe is homogeneous, isotropic and spatially
flat, thus it is  described by the metric,  
\begin{equation}\label{metric}
ds^2 = dt^2 - a^2(t)\; d\vec{x}^2\; .
\end{equation}

Anticipating the need for a non-perturbative treatment of the
evolution of the quantum state, we consider an inflaton 
model with an $N$-component scalar inflaton field $\vec{\Phi}(x) $
with quartic self-coupling. We then invoke the 
large $N$ limit as a non-perturbative tool to study the
dynamics\cite{noscos,asam,tsurob,tsu1,largeN}.  This choice 
is not only motivated by the necessity of a consistent
non-perturbative treatment but also because any 
grand unified field theory will contain a large number of scalar
fields, thus justifying a large $N$ limit on physical grounds.  

The matter action and Lagrangian density are given by
\begin{equation}
S [\vec\Phi] =  \int d^4x\; {\cal L}_m = \int d^4x \;
a^3(t)\left[\frac{1}{2} \, \left(\dot{\vec{\Phi}}\right)^2(x)-\frac{1}{2} \, 
\frac{\left(\vec{\nabla}\vec{\Phi}(x)\right)^2}{a^2(t)}-V(\vec{\Phi}(x))\right]\; ,
\label{action}
\end{equation}
\begin{equation}
V(\vec{\Phi})  =  \frac{m^2}2\; \vec{\Phi}^2 +
\frac{\lambda}{8N}\left(\vec{\Phi}^2\right)^2
+\frac12 \, \xi\; {\cal R} \;\vec{\Phi}^2  \;, \label{potential}
\end{equation}
and we will consider $ m^2 > 0 $, postponing the discussion of the
 case $ m^2 < 0 $ to future work. 
 Here $ {\cal R}(t) $ stands for the scalar curvature
\begin{equation}
{\cal R}(t)  =  6\left(\frac{\ddot{a}(t)}{a(t)}+
\frac{\dot{a}^2(t)}{a^2(t)}\right)\; , \label{ricciscalar}
\end{equation}
The $\xi$-coupling of $ {\vec\Phi}^2(x) $ to the scalar curvature $ {\cal
R}(t) $ has been included in the Lagrangian since it is necessary for
the renormalizability of the theory.

The discussion of the alternative inflationary mechanism that we are proposing and the physical description of the
quantum states  becomes more clear in
conformal time
\begin{equation} \label{conftdef}
{\cal T} = \int^t\frac{dt'}{a(t')}
\end{equation}
\noindent in terms of which  the metric is conformal to that in
Minkowski space-time  
\begin{equation}
ds^2 = a^2({\cal T})\,\left(d{\cal T}^2-d{\bf x}^2\right) \; .
\end{equation}
We introduce the conformally rescaled field 
\begin{equation}
\vec\Psi({\cal T}, {\bf x}) = a(t)\,\vec\Phi(t, {\bf x})\label{fieldrescale}
\end{equation}
\noindent in terms of which the  matter action becomes
\begin{equation}
{\cal S}[\vec\Psi] = \int{d{\cal T}\,d^3x\,
\left\{ \frac12\,[(\partial_{{\cal T}}\vec\Psi)^2-
(\nabla\vec\Psi)^2]-a^4({\cal T})\,V\left[\frac{\vec\Psi}{a({\cal T})}\right]+
a^2({\cal T})\,\frac{\cal R}{12}\,\vec\Psi^2 \right\}}\; .
\end{equation}
Since we are interested in describing the time evolution of an initial
quantum state, we pass on to the Hamiltonian description in the
Schr\"odinger representation. This procedure begins by obtaining 
the canonical momentum conjugate to the quantum field, 
$ \vec\Pi({\cal T}, {\bf x}) $, and the Hamiltonian density 
$ {\cal H}({\cal T}, {\bf x}) $
\begin{eqnarray} \label{confhamil}
\vec\Pi({\cal T},{\bf x}) &=& \vec\Psi'({\cal T},{\bf x})\; ,\cr\cr
{\cal H}({\cal T}, {\bf x}) &=& \frac12 \vec\Pi^2
  + \frac12\,(\nabla\vec\Psi)^2 
  + a^4({\cal T})\,V\left[\frac{\vec\Psi}{a({\cal T})}\right]
  - a^2({\cal T})\,\frac{\cal R}{12}\,\vec\Psi^2     \; ,     \cr\cr
H({\cal T}) &=& \int{d^3{\bf x} \; {\cal H}({\cal T}, {\bf x}) }\; ,
\end{eqnarray}
where the prime denotes derivative with respect to the conformal time 
$ {\cal T} $.

In the Schr\"odinger representation the canonical momentum is given by
\begin{equation}
\Pi^a({\cal T},{\bf x}) = - i \; \frac{\delta\;}
{\delta\Psi^a({\cal T},{\bf x})} \quad ;
\quad a = 1, \ldots, N\; .
\end{equation}
The time evolution of the wave-functional $
\Upsilon\left[\vec{\Psi};{\cal T} \right] $  is obtained from the
functional Schr\"odinger equation 
\begin{equation}
i\frac{\partial}{\partial {\cal T}} \Upsilon\left[\vec{\Psi};{\cal T}
\right] = H\left[\frac{\partial}{\partial
\vec{\Psi}};\vec{\Psi}\right] \Upsilon\left[\vec{\Psi};{\cal T}
\right] \label{schroedeqn} 
\end{equation} 
The implementation of the large $N$ limit begins by writing the field
as follows 
\begin{eqnarray} \label{fielddescomp}
\vec\Psi({\bf x},\,{\cal T}) 
&=& (\sigma({\bf x},\,{\cal T}),\,\vec\pi({\bf x},\,{\cal T}) )\cr\cr
&=& (\sqrt{N}\psi( {\cal T} ) + \chi({\bf x},\,{\cal T}) ,
\,\vec\pi({\bf x},\,{\cal T}) ) \; ,
\end{eqnarray}
\noindent where we choose the `1'-axis in the direction of  the
expectation value of the field and we collectively denote by
$\vec{\pi}$ the $N-1$ perpendicular directions.
\bea
\psi({\cal T}) & = &  \langle\sigma({\bf x},\,{\cal T})\rangle \nonumber \\
\langle\vec\pi({\bf x},\,{\cal T})\rangle & = &  \langle\chi({\bf
x},\,{\cal T})\rangle = 0 \label{fieldsplit} \; ,
\eea
\noindent where the expectations value above are obtained in the state
represented by the wave-functional 
 $ \Upsilon\left[\vec{\Psi};{\cal T} \right] $ introduced above. 

 The leading order in the large $N$ limit can be efficiently obtained by
 functional methods (see refs.\cite{noscos,asam,tsu1,tsu2,largeN} and
 references therein). The contributions of $ \chi $ to the equations
 of motion are subleading (of order $1/N$) in the large $ N $
 limit\cite{noscos,largeN}. 

It is convenient to introduce the spatial Fourier modes of the quantum  field 
\begin{equation}
\vec\pi_{{k}}({\cal T}) = \int{d^3x\;\vec\pi({\bf x},\,
{\cal T}) \; e^{i{\bf k}\cdot{\bf x}}}
\end{equation}
\noindent In leading order in the large $N$ limit, the explicit form
of the Hamiltonian is given by\cite{noscos,asam,tsu1,tsu2,largeN}   
\begin{eqnarray} \label{hamk}
H({\cal T}) &=& N\, {\cal V}\, h_{cl}({\cal T}) 
  - \frac{\lambda}{8\,N}\; \left(\sum_{k}
  \langle\vec\pi_{{k}}\cdot\vec\pi_{-{k}}\rangle\right)^2
  + \sum_{k} { H}_{k}({\cal T})\; ,  \cr\cr
h_{cl}({\cal T}) &=& 
\frac12\,\psi^{'2}({\cal T})+ \frac{a^2({\cal T})}2\; m^2 \; \psi^2({\cal T}) +
\frac{\lambda}{8}\,\psi^4({\cal T})  ,
\cr\cr
{ H}_{k}({\cal T})  &\equiv&  -\frac{1}{2}\;\frac{\delta^2\;\;}
  {\delta\vec\pi_{{k}}\cdot\delta\vec\pi_{-{k}}}
  + \frac{1}{2} \; \omega^2_k({\cal T})\;
\vec\pi_{{k}}\cdot\vec\pi_{-{k}} \label{hachek} \\ \cr
 \omega^2_k({\cal T}) &\equiv& {k^2} + a^2({\cal T}) \left[{\cal M}^2({\cal
T})-\frac{{\cal R}({\cal T})}{6} \right] 
\label{freq2} \; ,\\
{\cal M}^2({\cal T}) &\equiv& m^2 + \xi\,{\cal R} 
  + \frac{\lambda}{2}\,\frac{\psi^2}{a^2({\cal T})}
 +\frac{\lambda}{2}\,\frac{\langle\vec\pi^2\rangle}{N\,a^2({\cal
  T})}\label{M2} \; . 
\end{eqnarray}
where $ {\cal V} $ is the comoving volume. We assume spherically
symmetric distributions in momentum space. 

That is, in the large $N$ limit the Hamiltonian operator (\ref{confhamil})
becomes a time dependent c-number contribution plus a quantum mechanical 
contribution, $ \sum_{k} { H}_{k}({\cal T}) $, given by a collection
of harmonic oscillators with time-dependent frequencies, coupled only through
the quantum fluctuations $\langle\vec\pi_{{k}}\cdot\vec\pi_{-{k}}\rangle$.

In eqs.(\ref{hachek})-(\ref{M2}) the scale factor $ a({\cal T}) $ is
determined self-consistently by the Einstein-Friedmann equations. 

\subsection{Tsunami initial states} 

To highlight the main aspects of the inflationary scenario proposed
here, and to establish a clear difference with 
the conventional models, we first focus our discussion on  the case of
vanishing expectation value, i.e, $\psi({\cal T})=0$,  
and a {\em pure quantum state}. The most general
cases with mixed states described by density matrices and
non-vanishing expectation value of the field are 
discussed in detail in sections IIIC and  IV and in the appendix. 

For $\psi({\cal T})=0$ the quantum Hamiltonian (\ref{hamk}) becomes a
sum of harmonic oscillators with time dependent frequencies that
depend on the quantum fluctuations. Therefore, we propose a Gaussian
wave-functional of the form  
\be
\Upsilon\left[\vec{\Psi};{\cal T} \right] = {\cal N}_{\Upsilon}({\cal T})
  \prod_{k}\; e^{-\frac{A_{k}({\cal T})}{2}\;
  \vec\pi_{{k}}\cdot\vec\pi_{-{k}} } \label{gausswave} \; .
\ee
The functional Schr\"odinger equation (\ref{schroedeqn}) in this case
leads to evolution equations for the 
normalization factor ${\cal N}_{\Upsilon}({\cal T})$ and the
covariance kernel $ A_k({\cal T})$ whose general 
form is found in the appendix (see also \cite{noscos}). The evolution
of the normalization factor is determined by that of $ A_k$, while 
the equation for $ A_k$ is 
\be
i A'_k({\cal T}) = A^2_k -  \omega^2_k({\cal T}) \label{adoteqn}\; ,
\ee
\noindent where primes refer to derivatives with respect to conformal
time. As described in the appendix for 
the general case, the above equation can be linearized
by defining (see appendix)
\begin{equation} \label{defphik2}
 A_{ k}({\cal T}) \equiv - i\,
  \frac{\varphi^{'*}_{k}({\cal T})}{\varphi^*_{k}({\cal T})} \; ,
\end{equation}
where  the mode functions $ \varphi_{k} $ satisfy the equation
\begin{equation} \label{eqmod}
\varphi^{''}_{k} +  \omega^2_ k({\cal T})\; \varphi_{k}= 0 
\end{equation}
In terms of these mode functions the self-consistent expectation value
\begin{eqnarray}
\frac{\langle\vec\pi^2\rangle}{N} &=& \frac{1}{N}\int{\frac{d^3k}{(2\pi)^3}\,
  \langle\vec\pi_{{k}}\cdot\vec\pi_{-{k}}\rangle}
\cr\cr \langle\vec\pi_{{k}}\cdot\vec\pi_{-{k}}\rangle 
&=& \frac{N}{2A_{R, k}} = {N \over 2 } \; |\varphi_{k}|^2 \; .\label{selfcons}
\end{eqnarray}
We now must provide initial conditions on the wave functional to
completely specify the dynamics. Choosing the initial 
(conformal) time at ${\cal T}=0$ with $a({\cal T}=0)=1$, the initial
state is completely specified by furnishing 
the real and imaginary parts of the covariance $A_k$ at the initial
time. We parameterize these as\footnote{In the case for which
$ \omega^2_k(0)<0$ we choose $\omega_k(0)=\sqrt{k^2+|{\cal M}^2(0)-{\cal
R}(0)/6|}$.  } 
\be
A_{R,k}(0) = \Omega_k ~~; ~~ A_{I,k}(0)= \omega_k(0) \; \delta_k
\label{covariance} 
\ee
Choosing the Wronskian of the mode functions $ \varphi_{k}({\cal T}) $ and its
complex conjugate to be
\be
 \varphi_{k}\,\varphi^{'*}_{k} - \varphi^{'}_{k}\,\varphi^*_{k} = 2 i
\label{wronski} 
\ee
\noindent determines the following initial condition on the mode
functions (see appendix) 
\be
\varphi_{k}(0) = \frac{1}{\sqrt{\Omega_k}} ~~;~~ \varphi'_{k}(0)=
-\left[\omega_k(0)\delta_k +i \; \Omega_k\right]\varphi_{k}(0)\; .
\label{inicondmodes} 
\ee
where we also used eqs.(\ref{defphik2}) and (\ref{covariance}).

An important alternative interpretation of these mode functions is
that they form a basis for expanding the 
Heisenberg field operators (solution of the Heisenberg equations of motion)
\be
\vec{\pi}(\vec x,{\cal T}) = \int \frac{d^3 k}{(2\pi)^3} \left[
\vec{a}_{k} \; \varphi_{k}({\cal T}) \;e^{i\vec k \cdot \vec x}+
\vec{a}^{\dagger}_{k} \; \varphi^*_{k}({\cal T}) \; e^{-i\vec k \cdot \vec
x}\right] \label{heisenberg} \; ,
\ee
\noindent with $\vec{a}_{k};\vec{a}^{\dagger}_{k}$ annihilation and
creation operators, respectively, with 
canonical commutation relations. The Wronskian condition
(\ref{wronski}) ensures that the $ \vec{\pi}(\vec x,{\cal T}) $
fields and their conjugate momenta 
obey the canonical commutation relations at equal conformal times.  

The physical interpretation of these initial states is highlighted by
focusing on the occupation number 
of adiabatic states as well as on the probability distribution of
field configurations.  
\begin{itemize}
\item{{\em Occupation number:} it is at this point where the
description in terms of conformal time proves to be 
valuable. In conformal time the Hamiltonian in the large $N$ limit is
that of a collection of  harmonic 
oscillators with time dependent frequencies. It is then convenient to
introduce the adiabatic occupation number operator 
\be
\hat{n}_{k}({\cal T}) = \frac{1}{N}\left[\frac{{ H}_k({\cal T})}{
\omega_k({\cal T})}-\frac{1}{2}\right] \label{ocunumb}  \; ,
\ee 
\noindent with $H_k$ given by eq. (\ref{hachek}). 
 
In particular the occupation number at the initial time is given by
(see appendix) 
\be
n_k \equiv \langle \hat{n}_{k}(0) \rangle = \frac{\left[\omega_k(0)-\Omega_k
\right]^2+\omega^2_k(0) \;\delta^2_k}{4 \;\omega_k(0) \;\Omega_k}
\label{ocuini} \; . 
\ee
Here, the special case with $\Omega_k= \omega_k(0) $ and $ \delta_k=
0$ corresponds to the adiabatic vacuum ($ n_k = 0 $ ). Instead, 
we study an initial state in which a band of wave-vectors are
{\em populated with a non-perturbatively large number of
particles}. More precisely,  we consider initial states for which 
\bea
n_k &=& {\cal O}\left(\frac{1}{\lambda}\right) \quad   
\mbox{inside ~ the ~ excited ~ band}, \cr \cr
n_k &=&   0     \quad   \mbox{outside ~ the ~ excited ~ band.} \; .
\label{band}
\eea
\noindent where $\lambda$ is the quartic self-coupling.

This is accomplished by choosing,
\bea
&& {1 \over \Omega_k} ={\cal O}\left( {1 \over \lambda \; \omega_k(0)}\right)
\quad   \mbox{inside ~ the ~ excited ~ band,} \cr \cr
&&{1 \over \Omega_k} ={1 \over \omega_k(0)}\quad \mbox{and}\quad 
\delta_k = 0\quad \mbox{outside ~ the ~ excited ~ band.}\label{bigomega}
\label{lildelta}
\eea
These initial states are highly excited, the expectation
value of the energy momentum tensor in these states leads to an energy
density $\sim 1/\lambda$ and are, therefore, non-perturbative. We
will refer to the case where the excited band is narrow as the
narrow tsunami. 

It must be stressed that the particle distribution $ n_k $ alone {\em
partially} determines the initial state. As we see from
eq.(\ref{inicondmodes}), the initial state is completely defined
specifying {\em two} functions of $k : \;  \Omega_k$ and $ \delta_k $ . } 

\item{{\em Probability distribution:} an alternative interpretation of
these initial states is obtained by focusing on the 
probability distribution of field configurations at the initial
time. It is given by 
\be
{\cal P}[\vec{\pi}] = \left|\Upsilon\left[\vec{\Psi};{\cal T}=0
\right]\right|^2 = {\cal N}(0) 
  \prod_{k}\; e^{-\Omega_k\;
  \vec\pi_{{k}}\cdot\vec\pi_{-{k}} } \label{probability}
\ee
An intuitive quantum mechanical picture of the wave-functional for the
modes in the excited band is the following. At the 
initial time the instantaneous Hamiltonian corresponds to a set of
harmonic oscillators of frequencies $\omega_k(0)$, 
while the width in field space of the initial Gaussian state is
determined by $\Omega^{-1/2}_k$.
For a mode in the vacuum state $
1/\sqrt{\Omega_k} \sim 1/\sqrt{\omega_k(0)} $ and the typical
amplitudes of the field are $ {\vec \pi}_{ k} \sim 1/\sqrt{\omega_k(0)}
$ which is the typical width of the potential well. While for a mode
inside the excited band the width in field space 
is $ 1/\sqrt{\Omega_k} \sim 1/\sqrt{\lambda \; \omega_k(0)} $ [see
eq.(\ref{bigomega})]. Thus, 
large amplitude field configurations with $ {\vec \pi}_{ k\approx k_0}\sim
1/\sqrt{\lambda\; \omega_k(0)} \gg 1/\sqrt{\omega_k(0)} $ have
a probability of ${\cal O}(1)$, i.e, large amplitude configurations
within the band of excited wave-vectors are {\em not} suppressed. 
That is, the width of the probability distribution for these modes is
much larger than the typical size of the potential well and there is a
non-negligible probability for finding field configurations with large
amplitudes of ${\cal O}(1/\lambda)$. 

These highly excited initial states had been previously proposed as
models to describe the initial stages of ultrarelativistic heavy ion
collisions and had been coined `tsunami
waves'\cite{tsurob,tsu1,tsu2}. They represent spherical shells 
(in momentum space) with large occupation numbers of quanta,
describing a state with a large energy density with 
particles of a given momentum.  }
\end{itemize}

To highlight the difference between these states and other proposals
for inflation, we now summarize the important 
aspects of these tsunami-wave states before we study the
conditions under which such states  lead to  
inflation.

\subsubsection{Properties of Tsunami states}

\begin{itemize}
\item{{\em  Pure states:} the states under consideration, defined by
the wave-functional of the form (\ref{gausswave}) 
are {\em pure states} unlike some other proposals that invoke a
thermal distribution of particles, hence a mixed state 
density matrix as initial state.  }

\item{{\em  Vanishing expectation value of the scalar field:} Unlike
most models of classical chaotic inflation in which the scalar field 
obtains an expectation value, taken as a classical field, the
expectation value of the scalar field in the 
tsunami-wave states given by the wave-functional (\ref{gausswave})
{\em vanishes}. We also consider tsunami states with a non-zero
expectation value for the scalar field [see sec. IV]. } 

\item{{\em Highly excited initial modes:} the tsunami-wave
state described by eq. (\ref{gausswave}) with 
the covariance kernel given by eqs. (\ref{covariance}), (\ref{bigomega}) 
describes a  state for which the modes {\em inside} the excited band are
occupied with a non-perturbatively large ${\cal O}(1/\lambda)$  number
of (adiabatic) quanta. 
We remark that very high energy modes, 
those that will become superhorizon during the last 10 or so e-folds,
hence are of cosmological importance today, must be in the vacuum
state so as not to lead to  
a large amplitude of scalar density
perturbations\cite{revius,infl,noscos}.  } 
\end{itemize}
This type of {\em quantum states} is clearly a novel concept, it
presents an alternative to typical inflationary scenarios that 
invoke the dynamics of a {\em classical} scalar field which in most
cases ignore or bypass the quantum dynamics.   

The tsunami-wave states described above are the simplest states and
will be the focus of our study. These 
states can be generalized to describe mixed-state density matrices and
to also allow for an expectation value of the 
scalar field. These generalizations  are described in sections III.C
and IV and in the appendix and are found to lead  qualitatively the
same features  revealed by the simpler pure states.  

\subsection{Back to comoving time: renormalized equations of motion} 

Having set up the initial value problem in terms of the tsunami-wave
initial wave-functionals, the dynamics is now 
completely determined by the set of mode equations eqs.(\ref{eqmod}) with
 eqs.(\ref{freq2})-(\ref{M2}) and the initial 
conditions eqs.(\ref{inicondmodes}). It is convenient to re-write the
equations of motion in comoving time. This is accomplished by the
field rescaling  
given by eq.(\ref{fieldrescale}) which at the level of mode function
results in introducing the comoving time mode functions 
$f_k(t)$ related to the conformal time ones $\varphi_k({\cal T})$ as 
\be
f_k(t) = \frac{\varphi_k({\cal T})}{a(t)} \label{comomodes}
\ee
The equations of motion in comoving time for these mode functions are 
\begin{eqnarray}
&&\ddot f_{ k}(t) + 3\,H(t)\,\dot f_{ k}(t) + \left[
  \frac{k^2}{a^2(t)} + {\cal M}^2(t)  
  \right] f_{ k}(t) = 0 \label{comoeqn}\\
&&{\cal M}^2(t) = m^2 + \xi\,{\cal R}(t)  +
  \frac{\lambda}{4}\,\int\frac{d^3k}{(2\pi)^3}\,  |f_k(t)|^2
  \label{M2como}\\ 
&& f_{ k}(0) =  \frac{1}{\sqrt{\Omega_{k}}} \;; \quad \quad 
\dot f_{ k}(0) =  - [\omega_k(0)\,\delta_{ k} + H(0)
+ i \Omega_{ k}] \; f_{ k}(0) \label{inicondcomo} \; .
\end{eqnarray}
The Einstein-Friedmann equation are,
\be
H^2(t)=\left(\frac{\dot{a}(t)}{a(t)}\right)^2 = \frac{8 \pi
\rho_0}{3 M^2_{Pl}} \quad , \quad \rho_0= \langle T_{00}\rangle
\label{einstein} \; ,
\ee
\noindent where the expectation value is taken in the time evolved
quantum state.  
 It is straightforward to see that the expectation value of the
energy-momentum tensor   has the perfect fluid
form, as a consequence of the homogeneity and isotropy of the
system\cite{noscos,asam}. 

Thus the set of equations (\ref{comoeqn})-(\ref{einstein}) provide a
closed set of self-consistent equation 
for the dynamics of the quantum state and the space-time metric.

\subsubsection{Renormalized Equations of Motion in the Large $ N $ limit}

The  set of equations that determine the dynamics of the quantum state
{\em and} the scale factor 
need to be renormalized. The field quantum fluctuations
$$
\frac{\langle \vec \pi^2 \rangle}{N} = \int \frac{d^3k}{2(2\pi)^3}
\; |f_k(t)|^2 \; ,
$$
\noindent requires  subtractions which are absorbed in a
renormalization of the mass, coupling to the Ricci scalar and coupling
constant. The expectation value of the stress tensor also requires
subtractions (but not multiplicative renormalization). Since the
divergence structure is determined by the large energy, short distance
behavior, the 
band of excited modes does not influence the renormalization
aspects. Therefore, we use the extensive work on  
 the renormalization program which is  available in the literature
referring the reader to  references\cite{noscos,asam} 
for details. We here summarize the  aspects that are most relevant for
the present discussion.     

First, it is convenient to introduce the following dimensionless quantities,
\begin{eqnarray}
&&\tau = m \; t \quad ; \quad h(\tau)= \frac{H(t)}{m} \quad ; 
\quad q=\frac{k}{m} \quad ; 
\nonumber \\&& 
\omega_q = \frac{\omega_k}{m} \quad ; \quad
\Omega_q = \frac{\Omega_k}{m} \quad ; \quad g= \frac{\lambda}{8\pi^2}
\quad ; \quad  
f_q(\tau) = \sqrt{m} \; f_k(t) \; ,
\label{dimvars1} 
\eea
\noindent where $ m $ and $ \lambda $ stand for the renormalized mass of the
inflaton and the renormalized self-coupling, respectively\cite{noscos}.
In terms of these dimensionless quantities we now introduce the dimensionless
and fully renormalized expectation value of the self-consistent field as 
\bea
&& g\Sigma(\tau) \equiv  \frac{\lambda}{2m^2}\; \langle \pi^2(t)
\rangle_R \nonumber \\
&& \Sigma(\tau)= \int_0^{\infty} q^2 dq \left[ | f_q(\tau)|^2 - {1 \over
{q\; a(\tau)^2}} + {{\Theta(q - 1)}\over {2 q^3}} \left(\frac{{\cal
M}^2(\tau)}{m^2}-{{{\cal{R}(\tau)}}\over{6 m^2}}\right)\right] \;
.\label{sigre} 
\eea
\noindent where the terms subtracted inside the integrand renormalize
the mass, coupling to the Ricci scalar and
the coupling constant\cite{noscos,asam}. The dimensionless and renormalized
expressions for the energy density $\epsilon$ and pressure $p$ are given by
\bea
&&\epsilon(\tau)  \equiv {\lambda \over 2 N \; m^4} \langle
T^{00}\rangle_R =
\cr \cr
&&=     \frac{g\Sigma(\tau)}{2} + \frac{[g\Sigma(\tau)]^2}{4} +
\frac{g}{2}\int q^2 \; dq \left\{|\dot{f_q}(\tau)|^2 -
S_1(q,\tau)
+\frac{q^2}{a^2(\tau)} \left[|f_q(\tau)|^2 - S_2(q,\tau)\right]
\right\} \label{energydens} \cr \cr 
&& p(\tau)  \equiv {\lambda \over 2 N \; m^4} \, <T^{ii}>_R  \nonumber \\
&& (p+\epsilon)(\tau)   =  
g \int q^2 dq \left\{ |\dot{f_q}(\tau)|^2 - S_1(q,\tau)
+\frac{q^2}{3a^2(\tau)}\left[  |f_q(\tau)|^2 - S_2(q,\tau) \right] \right\}
\;. \label{pmase}
\end{eqnarray} 
Where the renormalization subtractions $ S_1 $ and  $ S_2 $ are given
by,\cite{noscos,asam}  
\begin{eqnarray}
S_1(q,\tau) &=&\frac{q}{a^4(\tau)}+\frac{1}{2qa^4(\tau)}
\left[B(\tau)+2\dot{a}^2  \right] \cr \cr   
&+& {\Theta(q - 1) \over {8 q^3 \; a^4(\tau) }}\left[ - B(\tau)^2 
- a(\tau)^2 {\ddot B}(\tau) + 3 a(\tau) {\dot a}(\tau) {\dot B}(\tau) 
- 4 {\dot a}^2(\tau) B(\tau) \right]\;,\cr\cr 
S_2(q,\tau) &=& \frac{1}{qa^2(\tau)}- \frac{1}{2q^3 a^2(\tau)}\;B(\tau)  
+ {\Theta(q - 1) \over {8 q^5 \; a^2(\tau) }}\left\{  3 B(\tau)^2 
+ a(\tau) \frac{d}{d\tau} \left[ a(\tau) {\dot B}(\tau)\right]\right\}\;,\cr\cr
B(\tau) &\equiv& a^2(\tau)\left[1+g\Sigma(\tau)\right] \; .\label{renosubs}
\end{eqnarray}
We choose  here $\xi =0$ (minimal coupling), the renormalization point 
$\kappa = |m|$ and $ a(0)=1 $.

In summary, the set of coupled, self-consistent 
equations of motion for the quantum state and the scale factor are
\begin{eqnarray}  
&& \left[\frac{d^2}{d \tau^2}+3h(\tau)
\frac{d}{d\tau}+\frac{q^2}{a^2(\tau)}+1+g\Sigma(\tau)
\right]f_q(\tau)  =  0 \label{modknr}  \\
&& f_q(0)  =  \frac{1}{\sqrt{\Omega_q}} \quad ; \quad
\dot{f}_q(0)  = - \left[\omega_q \; \delta_q + h(0)
+ i\,\Omega_q \right]f_q(0) \label{condini} \\
&&\omega_q = \sqrt{q^2+\left|1+g\Sigma(0)-\frac{{\cal
R}(0)}{6m^2}\right|}\label{omegaq} \; ,
\end{eqnarray}
plus  the Einstein-Friedmann equation of motion for the scale factor
\begin{equation} \label{h2tau}
h^2(\tau) = L^2 \, \epsilon(\tau) \qquad ,\qquad
\mbox{where } L^2 \equiv \frac{16 \, \pi N \, m^2}{3\, M^2_{Pl}\, \lambda} 
\; . \end{equation}
\noindent with $g\Sigma(\tau)$ and $\epsilon(\tau)$ given by
eqs. (\ref{sigre}) and (\ref{energydens}) respectively.  
 
In order to implement the numerical analysis of the set of 
eqs. (\ref{modknr})-(\ref{condini}), (\ref{sigre}) and (\ref{h2tau}) we
introduce an ultraviolet  momentum cutoff $ \Lambda $. For the cases
considered here we choose $ \Lambda \sim 200 $ and found
almost no dependence on the cutoff for larger values.  
As befits a scalar inflationary model,  the scalar self-coupling is
constrained by the amplitude of 
scalar density perturbations  to be $\lambda \sim
10^{-12}$\cite{revius,linde,lyth} implying that   $ g < 10^{-13} $. 
Therefore, the  subtractions
can be neglected because $ S_i \sim O(g\Lambda^4) < 10^{-4} $.

The initial state is defined by specifying the $\Omega_q$ and
$\delta_q$. We determine the range of these parameters $\Omega_q$ and
$\delta_q$ by the excitation spectrum for the tsunami-wave initial 
state, as well as the condition that lead to inflationary
stage. This will be studied in detail in the next section.  

\section{Tsunami inflation}

As emphasized in the previous section, the scenario under
consideration is very different from the popular treatments of
inflation based on the evolution of {\em classical } scalar inflaton 
 field\cite{cmb,revius,linde,lyth,infl}. In these scenarios all of the
initial energy is assumed to be in a zero mode (or 
order parameter) at the beginning of inflation and the quantum
fluctuations are taken to be perturbatively small with 
a negligible contribution to the energy density and the evolution of
the scale factor.  

In contrast to this description, our proposal highlights the dynamics
of the {\em quantum states} as the driving mechanism for inflation. 
The initial quantum states under consideration correspond to a band 
of quantum modes in excited states, thus the name
`tsunami-wave'\cite{tsurob,tsu1,tsu2}. This initial state  
models a cosmological initial condition in which the energy density
is non-perturbatively large, but concentrated in the  
quanta rather than in a zero mode.

We now study under which general conditions such a state can lead to a
period of inflation that satisfies the cosmological constraints for
solving the horizon and entropy problems entailing the necessity for about 
60 e-folds of inflation.    

It is understood that inflation takes place whenever the expansion of
the universe accelerates, i.e,
\be
 \frac{\ddot a}{a} = h^2 + \dot h = -\frac{L^2}{2}  [\epsilon+3p] > 0\; ,
\label{infla}  
\ee 
\noindent with $L$ given in eq. (\ref{h2tau}) and $\epsilon$ and $p$
given by eqs. (\ref{energydens}).

While our full analysis rely on the numerical integration of the above
set of equations, much we learn by considering the {\em narrow tsunami
case}.  

\subsection{Analytical study: the narrow tsunami case}
 
Before proceeding to a full numerical study of the equations of
motion, we want to obtain an 
analytic estimate of the conditions under which a tsunami initial
quantum state would lead to inflation.  

Our main criterion for such initial state to represent high energy 
excitations is that the number of quanta in the band of excited
modes is of ${\cal O}(1/\lambda)$. 
This criterion, as explained above, is tantamount to requiring that
field configurations with non-perturbative amplitudes 
have non-negligible functional probability. Progress can be made
analytically by focusing on the case in which the 
band of excited field modes is {\em narrow} i.e, its width $\Delta k $
is such that $\Delta k \ll  k_0$ or in terms of dimensionless quantities 
$\Delta q / q_0 \ll 1$. We introduce the following smooth distribution
\be
\Omega_q   =    \frac{\omega_q}{1+ \frac{{\cal N}_{\Omega}}{g}\;
e^{-\left[\frac{q-q_0}{\sqrt{2}\Delta q} \right]^2}} \quad , \quad
\mbox{with}\quad  \frac{\Delta q}{q_0} \ll 1\; ,
\label{distributions} 
\ee 
\noindent with $\omega_q$ given by eq. (\ref{omegaq}) and ${\cal
N}_{\Omega}$ a normalization constant that fixes the value of the
total energy. 

In addition, we choose $  \delta_q  = - h(0)/\omega_q $ as we
discuss below in eq.(\ref{slowrolldelta}).

This initial distribution posses the main features of the
tsunami state described in the previous section. 
Since $g \ll 1$, we have for $q \sim q_0$,
\be
\frac{1}{\Omega_q} \sim \frac{1}{g}\gg 1 ~~ \Rightarrow ~~ n_q \sim
\frac{1}{g} \gg 1 \label{biggy} 
\ee
\noindent corresponding to highly excited states. While for $|q-q_0|
\gg \Delta q$  
\be
\frac{1}{\Omega_q} \sim \frac{1}{\omega_q} ~~ \Rightarrow ~~ n_q \sim 0 \; .
\ee
\noindent Thus, these modes are in a  quantum state near the conformal
(adiabatic) vacuum at the initial time, with 
$n_q$ the number of quanta defined by eq. (\ref{ocuini}) in terms of
dimensionless variables. For these  
distributions (narrow tsunamis),  the integral over mode functions for
the quantum fluctuations $g\Sigma(\tau)$  [given by eq. (\ref{sigre})]
is dominated by the narrow band of excited states with mode amplitudes
$\sim 1/\sqrt{g}$ and can be approximated by 
\begin{equation} \label{apgsigI}
g\Sigma(\tau) = g\; \Delta q \; q_0^2\;  |f_{q_0}(\tau)|^2 + O(g)
+ O(g\,\Delta q) \simeq |\phi_{q_0}(\tau)|^2 \; ,
\end{equation}
where we have introduced the effective $q_0$ mode 
\be
 \phi_{q_0}(\tau) \equiv \sqrt{g \; \Delta q} \; q_0 \;
f_{q_0}(\tau). \label{effzeromode}
\ee
\noindent we note that the initial condition (\ref{condini})  and the
tsunami-wave condition (\ref{biggy}) entail that 
despite the presence of the coupling constant in its definition, the
amplitude of the effective $q_0$ mode is of ${\cal O}(1)$.  
 
The equation of motion for the effective $q_0$-mode takes the form
\begin{equation} \label{modoq0}
{\ddot \phi}_{q_0}(\tau) + 3 \, h(\tau) \, {\dot \phi}_{q_0}(\tau) + \left[
{q_0^2 \over a^2(\tau)} + 1 + |\phi_{q_0}(\tau)|^2
\right]\phi_{q_0}(\tau) = 0 \; .
\end{equation}
The scale factor follows from
\begin{equation} \label{friedq0}
h^2(\tau) = L^2 \; \epsilon(\tau) \; ,
\end{equation}
with energy and pressure,
\begin{eqnarray} \label{eypq0}
\epsilon(\tau) &=& \frac12 \, |{\dot \phi}_{q_0}(\tau)|^2 
+ \frac12 \, |\phi_{q_0}(\tau)|^2 + \frac14 \; |\phi_{q_0}(\tau)|^4 
+ \frac{q_0^2}{2 \, a^2(\tau)} \; |\phi_{q_0}(\tau)|^2 \; ,\cr \cr
(p+ \epsilon)(\tau) &=& |{\dot \phi}_{q_0}(\tau)|^2 + 
\frac{q_0^2}{3 \, a^2(\tau)} \; |\phi_{q_0}(\tau)|^2 \; .
\end{eqnarray}
\noindent where we have neglected terms of ${\cal O}(g)$. 

We will refer to the set of evolution equations
(\ref{modoq0})-(\ref{eypq0}) as the one mode approximation evolution equations.

In particular, within this one-mode approximation, the acceleration of
the scale factor obeys 
\begin{equation}
\frac{\ddot a(\tau)}{a(\tau)} = - L^2 \left[|{\dot \phi}_{q_0}(\tau)|^2 - \frac{|\phi_{q_0}(\tau)|^2}{2}-  \frac{|\phi_{q_0}(\tau)|^4}{4}  \right] \label{acc}
\end{equation}

Therefore, the condition for an inflationary epoch,  $ \ddot a > 0 $, becomes
\begin{equation}
|{\dot \phi}_{q_0}(\tau)|^2 \quad
< \quad \frac12 \, |\phi_{q_0}(\tau)|^2 + \frac14 \; |\phi_{q_0}(\tau)|^4 \;.
\end{equation}
A sufficient criterion that guarantees inflation is the {\em
tsunami slow roll condition}  
\be
 |{\dot \phi}_{q_0}(\tau)|  \ll
|\phi_{q_0}(\tau)| \label{slowroll}
\ee
The initial conditions (\ref{condini}) and the condition that
$\Omega_{q_0} \sim g \ll 1$ imply that the tsunami
slow roll condition (\ref{slowroll}) at early times is guaranteed if
$\delta_{q_0}$ is such that 
\be
|\omega_{q_0}\delta_{q_0} + h(0)| \ll 1 \label{slowrolldelta}
\ee
Hence tsunami-wave initial states that satisfy the tsunami
slow-roll condition (\ref{slowrolldelta}) {\em lead 
to an inflationary stage}. 

Moreover,  in order to have slow roll (\ref{slowroll}) at later times,
the effective friction coefficient $ 3 \, h(\tau) $  should be larger 
than the square of the frequency in the evolution equations
(\ref{modknr}). That is, 
\begin{equation}\label{rueda}
\frac{q_0^2+1+g\Sigma(0)}{3 h(0)} \ll 1 \;.
\end{equation}
(\emph{i.e.} the $q_0$-mode should be deep inside the overdamped 
oscillatory regime). Eq.(\ref{rueda})  implies that $ h(0) \gg 1 $ and
this together with eq.(\ref{slowrolldelta}) implies  that $ \delta_{q_0}
$ must be negative. 

A remarkable aspect of the narrow tsunami state is that it leads to a
dynamical evolution  
of the metric similar to that obtained in {\em classical chaotic inflationary
scenarios} in the slow roll approximation\cite{revius,linde,lyth}. In
particular the expression for the acceleration (\ref{acc}) and the
tsunami slow roll condition (\ref{slowroll}) are indeed similar to
those obtained in  classical chaotic inflationary models driven by a
homogeneous 
classical field (zero mode). However, despite the striking similarity with
 classical chaotic models, we haste to add that both the 
conditions that define a tsunami state and the tsunami slow roll
condition (\ref{slowroll}) guaranteed by the initial 
value (\ref{slowrolldelta}) is of purely quantum mechanical origin in
contrast with the classical chaotic-slow-roll 
scenario. Furthermore, we recall that the expectation value of the
scalar field vanishes in this state.  

\subsubsection{Early time dynamics} \label{tsuearly}

Under the assumption of a tsunami wave initial state 
and the tsunami slow-roll condition (\ref{slowroll})  
the contribution  $ {\dot \phi}_{q_0}(\tau) $  in the energy and
in the pressure [see eq. (\ref{eypq0})] can be neglected  provided,
\begin{equation} \label{estit}
|{\dot \phi}_{q_0}(\tau)|^2 \ll \frac{q_0^2}{3 \, a^2(\tau)} \;
 |\phi_{q_0}(\tau)|^2 \; .
\end{equation}
We call $ \tau_A $ the time scale at which this rely no longer holds. 
Furthermore, we can approximate $ \phi_{q_0}(\tau) $ by $
\phi_{q_0}(0) $ if
\begin{equation} \label{estit2}
\tau_A \ll \left| { \phi_{q_0}(0) \over {\dot \phi}_{q_0}(0)} \right| \; .
\end{equation}
This condition is fulfilled at least for $ \tau_A \lesssim 1 $ due to
the tsunami slow-roll condition (\ref{slowroll}).

During this interval the Friedmann equation (\ref{friedq0}) takes  the form,
\begin{equation} \label{friango}
\left[ {\dot a}(\tau) \over a(\tau) \right]^2 
=  L^2 \; \left[ \frac12 \, |\phi_{q_0}(0)|^2 + \frac14 \; |\phi_{q_0}(0)|^4 
+ \frac{q_0^2}{2 \, a^2(\tau)} \; |\phi_{q_0}(0)|^2 \right] 
= \frac{D}{a^2(\tau)} + E \; .
\end{equation}
where we used that $ g \Sigma(0) = |\phi_{q_0}(0)|^2 $. 
This equation is valid as long as the characteristic time scale of variation of the
metric  is  shorter than that  of the mode
$ \phi_{q_0}(\tau) $. 

The preceding equation can be integrated with solution
\begin{eqnarray} \label{soltaua}
a(\tau) &=& \sqrt{D \over E} \, \sinh\left(\sqrt{E}\, \tau
+c\right) \quad , \quad \frac{\ddot a(\tau)}{a(\tau)} =  E \;>\; 0 \;,
\cr\cr 
h(\tau) &=& \sqrt{E} \, \coth\left(\sqrt{E}\, \tau  + c\right)
 \quad , \quad
{\dot h}(\tau) = - { E \over \sinh^2\left(\sqrt{E}\, \tau
+c\right)} \; , \label{tsuni}  
\end{eqnarray}
where the constants $ D $, $ E $ and $ c $ are given by,
\begin{equation} \label{ctestaua}
D = L^2 \; \frac{q_0^2}{2} \; g\Sigma_0 \quad , \quad
E = L^2\;\left( \frac{g\Sigma_0}{2} + \frac{g\Sigma_0^2}{4}\right) \quad,\quad
\sinh c = \sqrt{\frac{E}{D}} \; ,
\end{equation} 
and $ g\Sigma_0 \equiv g \Sigma(0) = |\phi_{q_0}(0)|^2 $.

We see from eqs. (\ref{soltaua}) that during this interval there is  an
inflationary stage with an accelerated expansion $ \frac{\ddot
a(\tau)}{a(\tau)} = E > 0 $.  
We also see that $ h(\tau) $ decreases with time until it reaches the constant 
value $ \sqrt{E} $ that determines the onset of a quasi-De Sitter inflationary
stage. In ref.\cite{tsuinf} is shown that the solution in
eqs. (\ref{soltaua}) is valid for $ \tau < \tau_A $, where
\begin{equation} \label{taua}
\tau_A \sim {1 \over \sqrt{E}}\left[ \mbox{ArgSinh}
\left(\frac{\sqrt{3} \, q_0 \, E} 
{\sqrt{D}(1+g\Sigma_0)} \right)\;-\;c \right] 
= {1 \over \sqrt{E}}\left[ \mbox{ArgSinh} \left(
\frac{L\sqrt{3g\Sigma_0}\, (1+g\Sigma_0/2)}{\sqrt{2}(1+g\Sigma_0)}
\right)\;-\;c 
\right]\;. 
\end{equation}
This initial inflationary period with a  decreasing Hubble
parameter exists provided the r.h.s. is here positive, {\it i.e.},
\begin{equation} \label{q0condtaua}
q_0 > \frac{1+g\Sigma_0}{\sqrt{3\,E}} \; .
\end{equation}
In order to distinguish this phase from the later stages, to be
described below, we 
refer to this early time inflationary stage as `tsunami-wave
inflation' because the 
distinct evolution of the scale factor during this stage is
consequence of the tsunami-wave properties. 

At $ \tau = \tau_A $ we have:
\begin{eqnarray} \label{valuesattaua}
\phi_{q_0}(\tau_A) \simeq \phi_{q_0}(0) = \sqrt{g\Sigma_0} \qquad &,&  \qquad
\dot\phi_{q_0}(\tau_A) \simeq  - \frac{1+g\Sigma_0}{3\,h(\tau_A)} 
\; \phi_{q_0}(\tau_A) \;, \cr\cr
a(\tau_A) \simeq \frac{\sqrt{3\,E}\,q_0}{1+g\Sigma_0} \qquad &,&
\qquad h(\tau_A) \simeq  \sqrt{E} \; .
\end{eqnarray}

For $ \tau  >  \tau_A $, $ q_0/a(\tau) <<1 $ and the physical
wavevectors in the excited band have red-shifted so 
much that all  terms containing $ q_0 $ become negligible in the
evolution equations. Therefore, all modes in the excited band evolve
as an {\em effective}  $ q= 0 $ mode.  
Hence for $\tau > \tau_A$ the dynamics of the scale factor is
described by an effective homogeneous zero mode and 
describes a different regime from the one studied above. Such regime is akin to
the classical chaotic scenario.  

\subsubsection{The  effective classical chaotic inflationary
epoch}\label{tsulate} 

For $\tau >>\tau_A$, when $ q_0^2/a^2 \ll |\dot\phi_q|^2/|\phi_q|^2 \ll 1 $ 
 all the physical momenta corresponding to the comoving wavevectors in
 the excited band 
have redshifted to become negligible in the equations of motion. The
 dynamics is now  determined by the following  
set of equations for the effective zero mode and the scale factor,

\begin{eqnarray} \label{evoleqtaua}
&&{\ddot \phi}_{q_0}(\tau) + 3 \, h(\tau) \, {\dot \phi}_{q_0}(\tau) + 
\left[1 + |\phi_{q_0}(\tau)|^2\right]\phi_{q_0}(\tau) = 0 \; , \cr \cr
&&h^2(\tau) = L^2 \; \epsilon(\tau) \; ,
\end{eqnarray}
where the energy and  pressure are given by,
\begin{eqnarray} \label{petaua}
\epsilon(\tau) &=& \frac12 \, |{\dot \phi}_{q_0}(\tau)|^2 
+ \frac12 \, |\phi_{q_0}(\tau)|^2 + \frac14 \; |\phi_{q_0}(\tau)|^4 \; ,\cr \cr
(p+ \epsilon)(\tau) &=& |{\dot \phi}_{q_0}(\tau)|^2 \; .
\end{eqnarray}

The initial conditions on $\phi_{q_0}$ and ${\dot\phi}_{q_0}$ are
determined by their values at the time $\tau_A$, while the 
slow-roll condition (\ref{slowroll}) determines that the {\em
imaginary} parts of $ \phi_{q_0} $ and $ \dot\phi_{q_0} $ are
negligible.  

Therefore, after $ \tau_A $ the dynamic is identical to that of a classical homogeneous 
field (zero mode) 
\begin{equation}
\eta_{eff}(\tau) = \mbox{Re}[\phi_{q_0}(\tau)] \; ,
\end{equation}
that satisfies the equations of motion,
\begin{eqnarray} 
&&\ddot\eta_{eff} + 3\,h\,\dot\eta_{eff} +(1+\eta_{eff}^2)\,\eta_{eff} = 0\;,
\cr\cr
&&h^2(\tau) = L^2 \; \epsilon(\tau) \; . \label{classy}
\end{eqnarray}
with energy and pressure,
\begin{eqnarray} \label{peeff}
\epsilon(\tau) &=& \frac12 \, \dot\eta_{eff}^2 + \frac12 \, \eta_{eff}^2 
+\frac14 \; \eta_{eff}^4 \; ,\cr \cr
(p+ \epsilon)(\tau) &=& \dot\eta_{eff}^2 \; , \label{enermo}
\end{eqnarray}
and initial conditions [using eq. (\ref{valuesattaua})],
\begin{eqnarray} \label{magtaua}
\eta_{eff}(\tau_A) &=& \phi_{q_0}(\tau_A) = \phi_{q_0}(0) = \sqrt{g\Sigma_0}
\; , \cr \cr
\dot\eta_{eff}(\tau_A) &=& \dot\phi_{q_0}(\tau_A) = 
- \frac{1+g\Sigma_0}{3\,h(\tau_A)} \; \eta_{eff}(\tau_A) \; .
\end{eqnarray}
Where the value of $ \dot\eta_{eff}(\tau_A) $ is determined by 
the slow roll condition ($ \Rightarrow  \ddot\phi_{q_0}(\tau_A) \simeq 0 $), 
the evolution eq. (\ref{modoq0}) and $a(\tau_A)$
and $h(\tau_A)$ are given by eq. (\ref{valuesattaua}). 

When $ g\Sigma_0 \ll 1 $, the quadratic term in the potential dominates, and 
we can integrate the previous equations to obtain 
\begin{equation}
\eta_{eff}(\tau) = \eta_{eff}(\tau_A) - \frac{\sqrt{2}}{3L}(\tau-\tau_A) \;,
\quad \mbox{(for $ g\Sigma_0 \ll 1 $)} \;.
\end{equation}

This evolution is similar to that of classical chaotic inflationary
models\cite{revius,linde,lyth}. Therefore 
for $\tau > \tau_A$ when the physical momenta in the excited band have
redshifted so much that their contribution 
in the equations of motion of the quantum modes and the energy and
pressure become negligible, the evolution of 
the quantum modes and the metric is akin to a classical chaotic
inflationary scenario driven by a homogeneous 
c-number scalar field. This equivalence allows us to use  the results
obtained for classical chaotic inflation.
Thus, as the classical slow roll condition
($ |\dot\eta_{eff}| \ll |\eta_{eff}| $) holds, the evolution of the
effective scalar field is 
overdamped and 
the system enters a quasi-De Sitter inflationary epoch.
This inflationary period ends when the slowly decreasing Hubble parameter
becomes of the order of the inflaton mass, i.e,  $ 3\,h \sim 1 +
\eta_{eff}^2 $. 
At this stage the effective classical field exits the overdamped
regime and starts 
to oscillate, the slow roll condition no longer holds  and a matter dominated
epoch ($ |\dot\eta_{eff}| \sim |\eta_{eff}| \Longrightarrow p \sim 0
$) follows.  

\subsubsection{Number of e-folds}
An important cosmological quantity is the total number of e-folds
during inflation. As discussed above, there are two different
inflationary stages, the first one is determined by the 
equations (\ref{friango})-(\ref{tsuni}) and characterized by a rapid
fall-off of the Hubble parameter approaching a quasi-De Sitter
stage. This new stage has been referred to as the
tsunami-wave inflationary stage above to emphasize that the
dynamics is determined by the distinct characteristics of the
tsunami-wave initial stage.  

The second stage is described by an effective zero mode and
the evolution equations (\ref{classy}, \ref{peeff})  and is akin to
the chaotic inflationary stage driven 
by a classical homogeneous scalar field. The crossover between the two
regimes is determined by the time 
scale $\tau_A$ (in units of the inflaton mass) and given by
eqs.(\ref{taua}) at which the contribution from 
the term $q^2_0/a^2(\tau)$ to the equations of motion becomes
negligible. Therefore there are to distinct contributions 
to the total number of e-folds, which is given by  
\begin{equation} \label{nenarrow}
N_e(q_0, h(0)) = \log a(\tau_A) + N_e(0, h(\tau_A)) \; ,
\end{equation}
where $ a(\tau_A) $ is given by eq. (\ref{magtaua}) and 
$ N_e(0, h(\tau_A)) $ is just the number of e-folds for classical chaotic 
inflation with an initial Hubble parameter $ h(\tau_A) $.

We can express $ h(\tau_A) $ as a function of $ q_0 $ and $ h(0) $,
\begin{equation}
h(\tau_A) = L \; \sqrt{\frac{g\Sigma_0}{2}+\frac{(g\Sigma_0)^2}{4}}
= \sqrt{h^2(0)-L^2\;\frac{q_0^2}{2}\;g\Sigma_0} \;.
\end{equation}
The number of e-folds during the first stage, is given by
\be
\log a(\tau_A) \sim \log\left[\frac{\sqrt{3\,E}\,q_0}{1+g\Sigma_0}
\right] \label{efoldtsunami} 
\ee
The expression for the number of e-folds during the following, chaotic
inflationary stage simplifies when $ g\Sigma_0 \ll 1 $. 
In this case the quadratic term in the potential dominates, and we can
obtain simple analytical expressions 
\begin{equation} \label{lnane0narrow}
N_e(0, h(\tau_A)) = \frac{3L^2}{4} \; \eta_{eff}^2(\tau_A) 
= \frac{3L^2}{4} \; g\Sigma_0 
= \frac{3L^2}{2} \; \frac{\epsilon_0}{1+q_0^2}     \;,
\quad \quad \quad 
\mbox{(for $ g\Sigma_0 \ll 1 $)} \;.
\end{equation}
We see that the number of efolds grow when $ q_0 $ decreases at fixed
initial energy $ \epsilon_0 $. That is, we have more efolds when the energy is
concentrated at low momenta. 

\subsubsection{In summary} 

Before proceeding to a full numerical study of the evolution we 
summarize the main features of the dynamics gleaned from the narrow
tsunami case to compare with the numerical results. 

\begin{itemize}

\item{The conditions for tsunami-wave inflation are {\em i) } a band
of excited states centered at a momentum $k_0$ 
with a non-perturbatively large ${\cal
O}(1/g)$ number of quanta in this band, and {\em ii)} the tsunami
slow-roll condition eq.(\ref{slowroll}). These conditions are
guaranteed by the initial conditions   
on the mode functions given by eq.(\ref{condini}) with the tsunami-wave
distributions of the general form given by
eqs. (\ref{distributions}), (\ref{slowrolldelta}). }   

\item{
There are two successive inflationary periods. During the  first
one, described in sec. \ref{tsuearly}, the dynamics is completely
characterized by the distinct features of the 
tsunami-wave initial state,  the Hubble parameter falls off fast and
reaches an approximately  constant value $ \sqrt{E} $ that
characterizes the quasi-De Sitter epoch of 
inflation of the second period. The second stage, described in
sec. \ref{tsulate}  can be described in terms of an effective
classical zero mode and the 
evolution of this effective mode and that of the Hubble parameter are
akin to the standard chaotic inflationary scenario.}

\item{
The tsunami-wave initial state can be interpreted as a {\em
microscopic} justification of the classical chaotic scenario described by an 
effective classical zero mode of large amplitude. The amplitude of
this effective zero mode is {\em non-perturbative} as 
a consequence of the non-perturbative ${\cal O}(1/\lambda)$ number of
quanta in the narrow band of 
excited modes. Thus the initial value of the effective, classical zero
mode that describes the second, chaotic inflationary stage, is 
completely determined by the quantum initial state. 
}

\item{An important point from the perspective of structure formation
is that the band of excited wavevectors centered at $q_0$ either 
correspond to superhorizon modes initially, or all of the excited
modes cross the horizon during the first stage of inflation, i.e, 
during the tsunami stage. This is important because the chaotic second
stage of inflation which dominates during a longer period guarantees 
that the band of excited modes have become superhorizon well before
the last 10 e-folds of inflation and hence cannot affect the 
power spectrum of the temperature anisotropies in the CMB. The fact
that the tsunami-wave initial state is such that the very high 
energy modes (necessarily trans-Planckian) that cross the horizon
during the last 10 e-folds and are therefore of cosmological 
importance today are in their (conformal) vacuum state leads to the
usual results from chaotic inflation for the power spectrum 
of scalar density perturbations.   }

\end{itemize}

Although these conclusions are based on the narrow tsunami
case, we will see below that a full numerical integration 
of the self-consistent set of equations of motion confirms this picture. 

In sections \ref{sotherdist} and \ref{sgenchaoinf} we show how
this results can be easily extended to {\em more general particle 
distributions} and {\em more general initial states}.

\subsection{Numerical example}

To make contact with familiar models of inflation with an inflaton
 field with a mass  near the grand unification scale, we choose the 
 following  values of the parameters:  
\be 
 \frac{m}{ M_{Pl}} = 10^{-4}~ ,~
\; \lambda = 10^{-12} ~ , ~ \; N = 20 
\ee
\noindent where the number of scalar fields $N=20$ has been chosen as
a generic representative of a grand unified quantum field theory.  
 For these values we find 
$$
L^2 \equiv \frac{16 \, \pi N \, m^2}{3\, M^2_{Pl}\, \lambda} 
= 3.35 \cdot 10^6 \; .
$$

As an example we shall consider an initial energy density  
$ \rho_0 = \langle T_{00} \rangle  = 10^{-2} \, M_{Pl}^4 $.
Thus, the initial value for the Hubble parameter is 
$ H_0 = \sqrt{8 \pi  \rho_0 / 3M_{Pl}} = 3.53 \cdot 10^{18} \; GeV 
(= 1.654 \cdot 10^{52} \; km/s/Mpc) $.
These initial conditions in dimensionless  variables give 
$ \epsilon_0 = 2.50 $ and $ h(0) = 2890 $.

In addition, the slow roll conditions (\ref{rueda}) imply:
$$
\frac{q_0^2+1+g\Sigma_0}{3\,h(0)} \ll 1
$$
which  in this case results in
$$
q_0 \ll 95 \;.
$$
We choose $ q_0 = 80.0 $, and initial conditions in
 eq. (\ref{condini}) with  $\Omega_q$ and $\delta_q$ given by
 eq. (\ref{distributions}) and (\ref{slowrolldelta}). These initial conditions
 satisfy the tsunami slow roll condition,
\begin{eqnarray}
|\omega_q\;\delta_q+h(0)| \ll 1 
\end{eqnarray}
Furthermore,  we take $ \Delta q = 0.1 $ and $ {\cal N}_\Omega $ is
adjusted by fixing the value $ g\Sigma(0) = g\Sigma_0 $ which for the
values chosen  for $ \epsilon_0 $ and by
eqs.(\ref{energydens}) and (\ref{apgsigI}) and (\ref{slowroll}) gives
$ g\Sigma_0 = 7.81 \cdot 10^{-4} $.  

\begin{figure}[h]
\begin{turn}{-90}
\epsfig{file=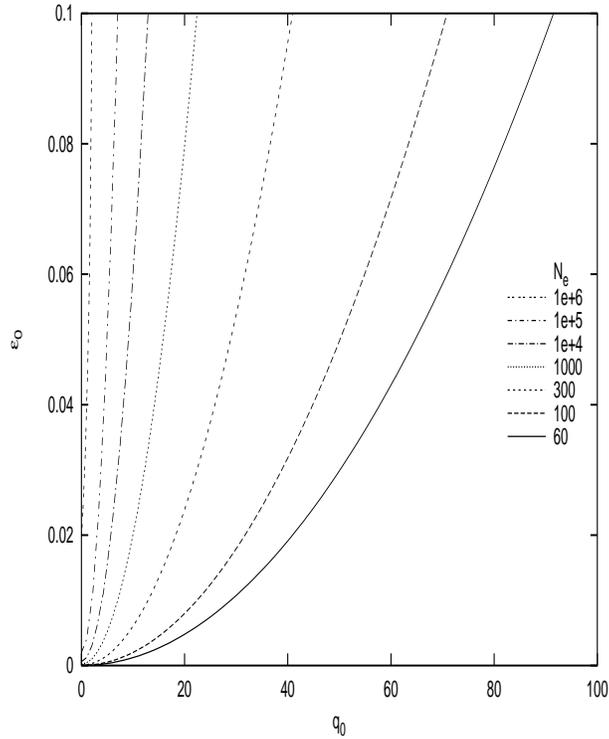,width=10cm,height=8cm}
\end{turn}
\vspace{.1in}
\caption{Tsunami  inflation: isolines of 
constant number of efolds obtained from eq.(\ref{lnane0narrow})
(valid for $ g\Sigma_0 \ll 1 $), 
for $ m = 10^{-4} M_{Pl} $, $ \lambda = 10^{-12} $ and $ N = 20 $.}
\label{Nefig}
\end{figure}

\begin{figure}[h]
\begin{turn}{-90}
\epsfig{file=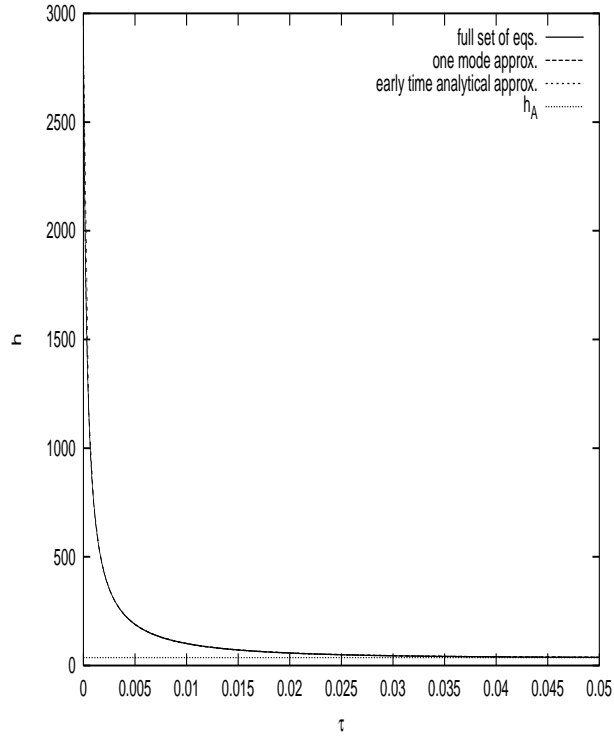,width=10cm,height=8cm}
\end{turn}
\vspace{.1in}
\caption{Tsunami  inflation: 
Early time $ h(\tau) $. $ h_A \equiv h(\tau_A)$ is the asymptotic value for
the early period that ends at $ \tau_A \sim 0.133 $.
For $ m = 10^{-4} M_{Planck} $, $ \lambda = 10^{-12} $ and $ N = 20 $.
Initial conditions: 
$ \rho_0 = 10^{-2} M_{Pl}^4 $, $ q_0 = 80.0 $ and $ \Delta q = 0.1 q_0 $.}
\label{hearlyfig}
\end{figure}

\begin{figure}[h]
\begin{turn}{-90}
\epsfig{file=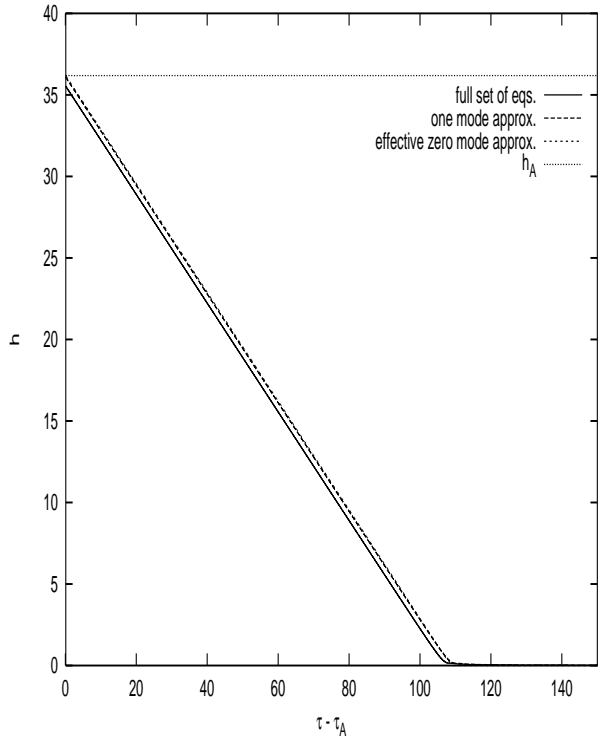,width=10cm,height=8cm}
\end{turn}
\vspace{.1in}
\caption{Tsunami  inflation: 
$ h(\tau) $ for $ \tau > \tau_A $. The early time analytic
approximation gives $ h_A = 36.1 $ (also with the one mode approx.), 
numerically we obtain $ h(\tau_A) = 35.5 $. 
Same parameters and initial conditions as in fig. \ref{hearlyfig}.} 
\label{hfig}
\end{figure}

\begin{figure}[h]
\begin{turn}{-90}
\epsfig{file=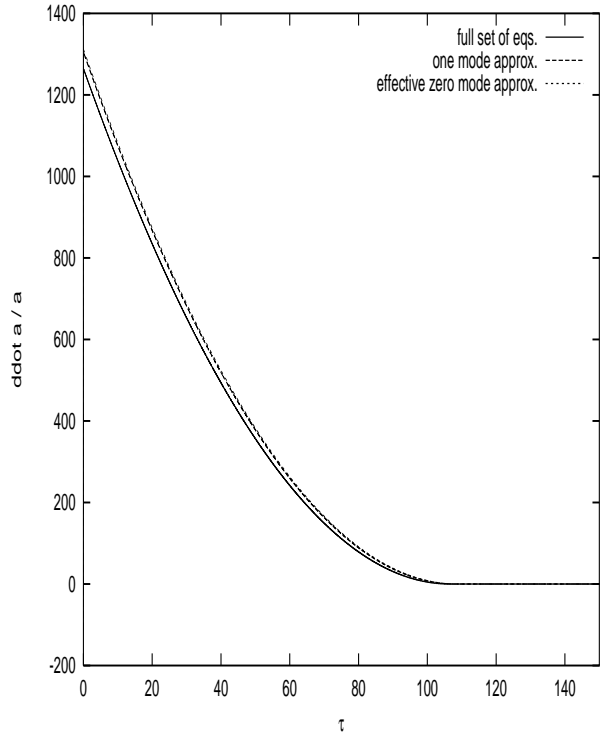,width=10cm,height=8cm}
\end{turn}
\vspace{.1in}
\caption{Tsunami  inflation:
$ \ddot a(\tau) \over a(\tau) $, it shows that there is accelerated 
expansion (inflation) up to times $ \tau \sim 109 $. 
Same parameters and initial conditions as in fig. \ref{hearlyfig}.}
\label{ddaoverafig}
\end{figure}

\begin{figure}[h]
\begin{turn}{-90}
\epsfig{file=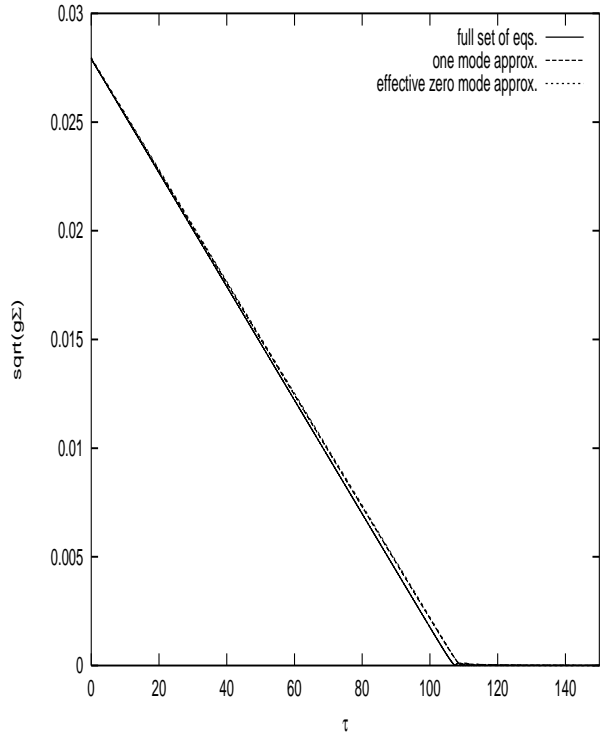,width=10cm,height=8cm}
\end{turn}
\vspace{.1in}
\caption{Tsunami  inflation: 
$ \sqrt{g\Sigma(\tau)} $, after $ \tau_A \sim 0.133 $, it plays 
the role of an effective classical field. 
Same parameters and initial conditions as in fig. \ref{hearlyfig}.}
\label{gsigmafig}
\end{figure}

\begin{figure}[h]
\begin{turn}{-90}
\epsfig{file=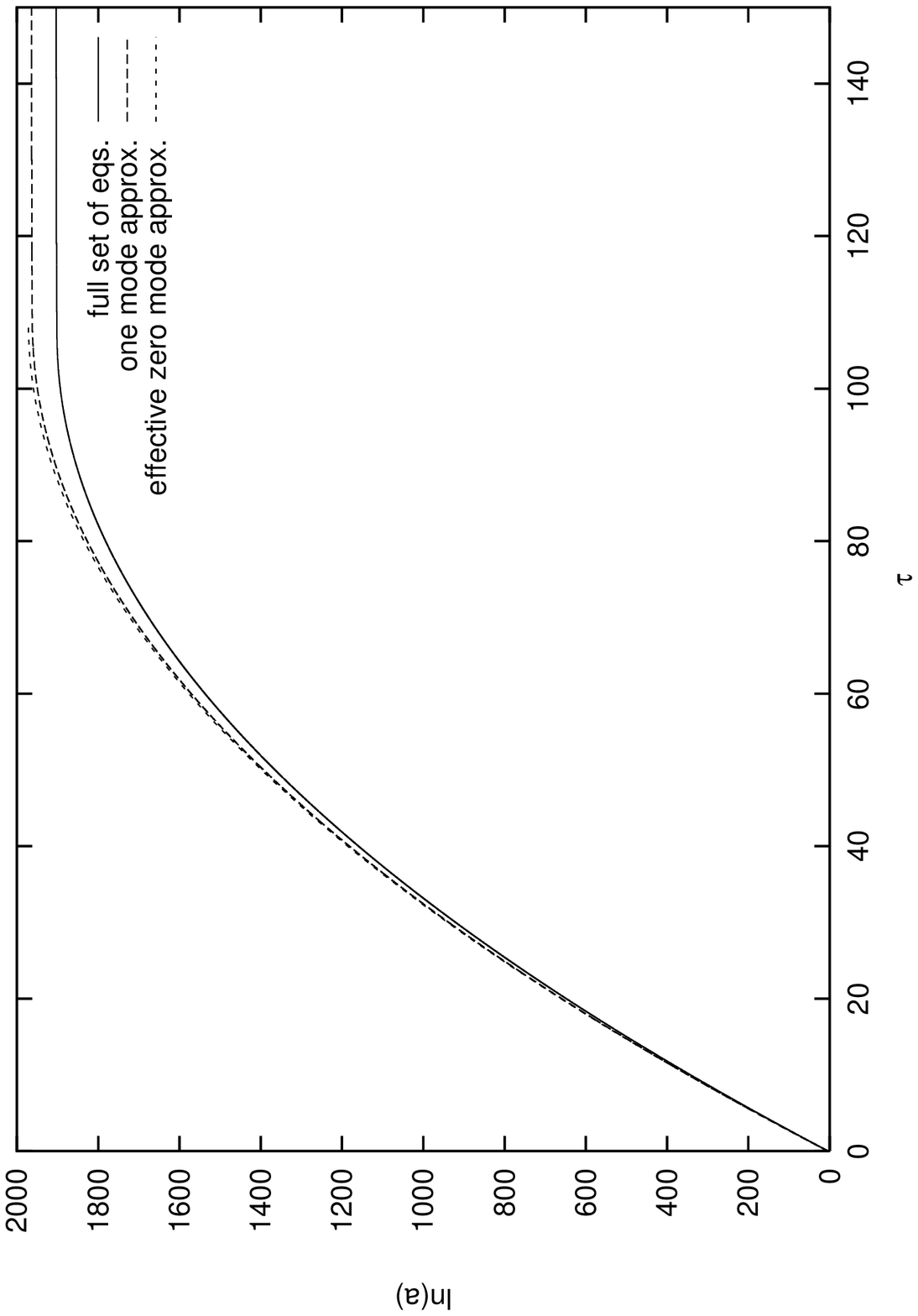,width=10cm,height=8cm}
\end{turn}
\vspace{.1in}
\caption{$ \ln[a(\tau)] $ vs. $\tau$.
Same parameters and initial conditions as in fig. \ref{hearlyfig}.}
\label{lnafig}
\end{figure}

\begin{figure}[h]
\begin{turn}{-90}
\epsfig{file=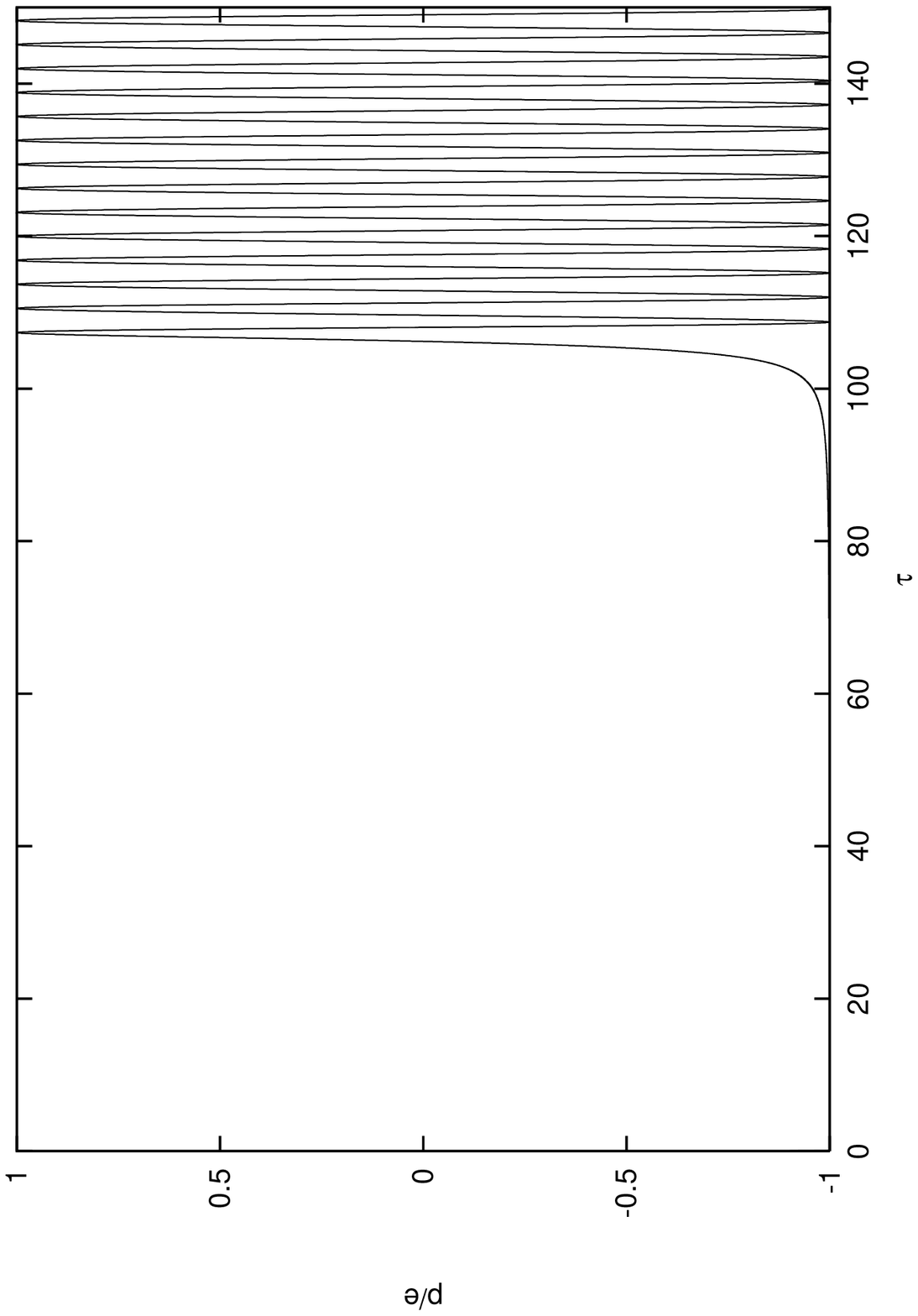,width=10cm,height=8cm}
\end{turn}
\vspace{.1in}
\caption{Tsunami  inflation: 
$ p(\tau)/\epsilon(\tau) $. It shows the onset of a matter dominated
epoch after the quasi-De Sitter stage.
Same parameters and initial conditions as in fig. \ref{hearlyfig}.}
\label{poverefig}
\end{figure}

Figure \ref{Nefig} displays $\epsilon_0$ vs $q_0$ along lines of
constant number of e-folds , while figures
\ref{hearlyfig}-\ref{poverefig} display the solution of the 
full set of equations  (\ref{modknr})-(\ref{h2tau})
with (\ref{energydens}). An important feature that
emerges from these figures is that for the set of parameters that are typical
for inflationary scenarios and for large values of $q_0=k_0/m$ (but 
well below the Planck scale) the number of e-folds obtained is more
than sufficient as shown by fig.\ref{lnafig}.  

\vspace{1cm}

\vspace{1cm}

We also show that the dynamics of the full set of equations
(\ref{modknr})-(\ref{h2tau}) with
(\ref{energydens}) is correctly approximated  by the narrow tsunami
case studied in the previous subsections:  
the one mode approximation [eqs. (\ref{apgsigI})-(\ref{eypq0})],
the early time analytical formulae (for $ \tau \le \tau_A $) 
[eqs. (\ref{soltaua})-(\ref{ctestaua})],
and the effective classical field (for $ \tau > \tau_A $)
[eqs. (\ref{peeff})]. The agreement between the analytic treatment and
the full numerical evolution is displayed in figures  
\ref{hearlyfig}-\ref{poverefig}.

The early time analytic expressions predict an inflationary
period during which the Hubble parameter falls off fairly fast, that
lasts up to 
$ \tau_A \sim 0.133 $ [eq. (\ref{taua})] reaching an asymptotic value of 
$ h(\tau_A) = 36.2 $ [eqs. (\ref{ctestaua}) and (\ref{valuesattaua})]. 
The one mode approximation gives the same prediction $ h(\tau_A) = 36.1 $, 
and numerically evolving the full set of equations we find
$ h(\tau_A) =  35.5 $.
Thus, we see from this values and from figs. \ref{hearlyfig} and 
\ref{hfig} that both approximations are fairly accurate  for early
times.

After $ \tau_A $, the geometry reaches a quasi-De Sitter epoch. We
have shown in the  
previous subsection that after the time $ \tau_A $ the evolution
equations for the 
one mode approximation reduce to those of an effective classical field.
The effective zero mode approximation correctly predicts the dynamics
in this epoch as can be gleaned from figures \ref{hfig}-\ref{lnafig}.

While the stage of early tsunami inflation up to  $ \tau_A $ results in
only $ 8.5 $ efolds, the following quasi-de Sitter stage described by the 
effective classical scalar field 
lasts for  a total of $ 1900 $ efolds. For the values of 
parameters chosen above,  $ g\Sigma_0 \ll 1 $, hence we can estimate the number
of efolds with eq. (\ref{lnane0narrow}). Using eq. (\ref{nenarrow}) we obtain 
a total of $ 1970 $ efolds while 
the one mode approximation yields $ 1960 $ efolds. Both results agree
with the full numerical solution of the equations (see fig. \ref{lnafig}). 

Furthermore, as stated above  inflation ends when 
$ h \sim \frac{1+\eta_{eff}^2}{3} \sim \frac{1}{3} $, after which  a matter
dominated epoch follows.

\subsection{Other distributions and other states:} \label{sotherdist}
The validity of the physical picture that emerges from the previous analytic
and numerical study is not restricted to 
pure states or narrow distributions of the form given by
(\ref{distributions}). We have also studied more general 
distributions and mixed states:

\vspace{1mm}

{\em Other distributions: }

The narrow tsunami case where a single quantum mode $q_0$
dominates the dynamics has been extremely useful to study the dynamics in the
previous section. The generalization to the case with 
continuous distributions of $q$-modes can be easily obtained making
the changes:
\begin{eqnarray}\label{cambio}
|\phi_{q_0}(\tau)|^2 &\to& g \, \int q^2 \; dq \; |f_q(\tau)|^2 \; , \cr \cr
|{\dot\phi}_{q_0}(\tau)|^2 &\to& g \, \int q^2 \; dq \; |{\dot f}_q(\tau)|^2
\; , \cr \cr 
q_0^2 \; |\phi_{q_0}(\tau)|^2 &\to& g \, \int q^2 \; dq \; q^2
\; |f_q(\tau)|^2 \; .  
\end{eqnarray}
The two stages of inflation are always present for such continuous
modes distribution as long as the following generalized slow-roll
conditions is fulfilled
\begin{equation}\label{genslo}
g \, \int q^2 \; dq \; |{\dot f}_q(\tau)|^2 \ll
g \, \int q^2 \; dq \; |f_q(\tau)|^2 
\end{equation}
that imposes on $ \delta_q $ the condition $ |\omega_q \, \delta_q +
h(0)| \ll 1 $. 

The effective zero mode in the second stage of inflation is now given by
$$
{\eta}^2_{eff}(\tau) = g \, \int q^2 \; dq \; |f_q(\tau)|^2 \; .
$$
Our numerical study with general distributions reveals that the
analytical picture obtained by substituting eq.(\ref{cambio}) in
sec. IIIA correctly reproduce the dynamics.

\vspace{1mm}

{\em Other (mixed) states:} Although we have focused for simplicity on
tsunami pure initial states, we have also 
investigated the possibility of mixed states. Mixed state density
matrices and their time evolution are discussed in the appendix. The 
mixing can be parametrized in terms of angles $\Theta_k$ as given in
equation (\ref{defcalA}) and the number of (conformal)  quanta 
are given by eq. (\ref{occumix}). The {\em only} relevant changes that
occur are  in the integrals for
$\Sigma(\tau)~;~\varepsilon(\tau)~;~p(\tau)$ in which  
\be
|{f}_q(\tau)|^2 \rightarrow |{f}_q(\tau)|^2
\coth\left[\frac{\Theta_q}{2}\right]
 \quad , \quad
|\dot{f}_q(\tau)|^2 \rightarrow |\dot{f}_q(\tau)|^2
\coth\left[\frac{\Theta_q}{2}\right] \label{change} 
\ee
Tsunami and slow-roll conditions on the mode functions given by
eqs. (\ref{bigomega}) and (\ref{slowrolldelta}) lead
to tsunami-wave inflation followed by chaotic inflation just as
discussed above. In the narrow 
tsunami case the only change is that the effective $q_0$-mode
is rescaled by the mixing factor, i.e, 
\be
|\phi_{q_0}(\tau)|^2 \rightarrow |\phi_{q_0}(\tau)|^2
\coth\left[\frac{\Theta_{q_0}}{2}\right] \quad , \quad
|{\dot\phi}_{q_0}(\tau)|^2 \rightarrow |{\dot\phi}_{q_0}(\tau)|^2
\coth\left[\frac{\Theta_{q_0}}{2}\right]
\label{mixedq0modo} 
\ee 
It is also illuminating to contrast the tsunami-wave mixed
states with the more familiar {\em thermal} mixed states. The latter
are obtained by the choice  
\be
\Omega_q = \omega_q ~~; ~~ \delta_q = 0 ~~; ~~  \Theta_q
= \frac{\omega_q}{T} \label{thermal} 
\ee 
\noindent with $T$ some value of temperature. In this case it is
straightforward to see that (quantum) equipartition 
results in that the contributions of the modes and their time
derivatives to the energy and  pressure are of the same order  
  ($ |\dot{f}_q(\tau)|^2 \sim [ h(0)^2 +\omega_q^2] \;
|{f}_q(\tau)|^2$). Hence, for 
these thermal mixed states the tsunami slow-roll condition is not fulfilled.  
This is obviously {\em not} surprising, such a choice of thermal mixed
state leads to a FRW epoch which 
is not inflationary. Hence the tsunami-wave initial conditions along
with the generalized slow-roll conditions lead to {\em two} successive
inflationary epochs in striking contrast to the familiar mixed thermal
states.  
\section{Generalized chaotic inflation} \label{sgenchaoinf}

The previous analysis, confirmed by the numerical evolution of the
full self-consistent set of equations leads to one of the important
conclusions that tsunami-wave initial states provide
a microscopic justification of the chaotic inflationary scenario. 

We have focused our discussion on initial states with vanishing 
expectation value of the scalar field (order parameter) and where the
energy is concentrated in a momentum band (tsunami initial states). 
This choice brings to the fore
the striking contrast between this novel {\em quantum state} and the usual
classical approach to chaotic inflation. 

A relevant question to pose is what would be the dynamics if the initial state
allows for a non-vanishing expectation value of the scalar field {\em
along} with 
some of the initial energy localized in excited quanta. In this
section we address this 
question and study in detail this alternative scenario, which we refer
to as generalized 
chaotic inflation to distinguish from the tsunami-wave state studied
above. This 
generalization thus includes both cases: the classical chaotic inflation in
the limit when 
there are no excited modes, as well as  the tsunami initial state when all
of the energy 
is localized in a band of excited modes and the expectation value of
the field vanishes.  

The relevant equations of motion in comoving time for the mode
functions in this case are given by (\ref{equationsgene}) in the appendix.
Along with the dimensionless variables (\ref{dimvars1})  it is also 
convenient to introduce a dimensionless expectation value as
\be
\eta^2(\tau) = \frac{\lambda}{2m^2} \; \phi^2(t) \label{dimlessexp}
\ee

In this generalized case with $\eta \neq 0$ the equations of motion
for the mode functions $f_q(\tau)$  
(in terms of dimensionless variables) are the same as in
eq. (\ref{modknr}) after the replacement  
$g\Sigma(\tau)\rightarrow g\Sigma(\tau)+\eta^2(\tau)$ and the equation
of motion for $\eta(\tau)$ is given by 
\begin{eqnarray}
&& \frac{d^2 \eta({\tau})}{d\tau^2}+3h(\tau)
\frac{d\eta(\tau)}{d\tau}+\left[1+\eta^2(\tau)+g\Sigma(\tau)
\right]\eta(\tau)  =  0 \label{etacomo}  \\
&& \eta(0)  =  \eta_0 \quad ; \quad
\dot{\eta}(0)  = \dot{\eta}_0 \label{condinieta} 
\end{eqnarray}
The Einstein-Friedmann equation is given by (\ref{h2tau}) but with the
energy density and pressure now given by  
\begin{eqnarray}\label{enepre}
\epsilon(\tau) &  =  &  \frac{1}{2}\dot{\eta}^2+ \frac12 \,
\left(g\Sigma+\eta^2\right) + 
\frac14 \, \left(g\Sigma+\eta^2\right)^2 + \nonumber \\ 
&+ & \frac{g}{2}\int q^2 \; dq \left\{|\dot{f_q}|^2 -  
S_1(q,\tau) +\frac{q^2}{a^2}\, \left[|f_q|^2 - S_2(q,\tau)\right]
\right\} \; ,\label{enerdensgene}\\
 (p+\epsilon)(\tau)  & = & 
\dot{\eta}^2  +   g \int q^2\; dq \left\{ |\dot{f_q}|^2 - S_1(q,\tau)
+\frac{q^2}{3a^2}\left[  |f_q|^2 - S_2(q,\tau) \right] \right\}
\;. \label{pmasegene} 
\end{eqnarray}
\noindent where the renormalization subtractions $ S_1~;~S_2 $ are
obtained from those given by eq. (\ref{renosubs}) upon the 
replacement $g\Sigma \rightarrow g\Sigma+\eta^2$. 

From this expression we see that for  fixed (large) energy density
Status: RO

there are two different possibilities: if the  zero
mode squared $\eta^2(0)$ is larger than
the quantum fluctuations $ g\Sigma $ and $g \int q^4 \; dq \; |f_q|^2 $,
the dynamics is basically similar to that in the usual chaotic
inflationary scenarios. This 
corresponds to most of the initial energy density to be in the zero mode and 
little energy density in the band of excited states. On the other hand,
for small $\eta^2(0)$ most of the initial energy density is 
in the tsunami quantum state and the initial dynamics is akin to the
$\eta=0$ case. To quantify this statement and clarify the 
interplay and crossover of behaviors between the $\eta=0$ and the
generalized chaotic case, we now resort again to the narrow 
tsunami case, which highlights the essential physics. 

The relevant equations are: i) the equations of motion for the
effective $q_0-$mode (\ref{effzeromode}), 
\begin{equation} 
{\ddot \phi}_{q_0}(\tau) + 3 \, h(\tau) \, {\dot \phi}_{q_0}(\tau) + \left[
{q_0^2 \over a^2(\tau)} + 1 + \eta^2(\tau)+|\phi_{q_0}(\tau)|^2
\right]\phi_{q_0}(\tau) = 0 \; .\label{modoq0gen}
\end{equation}
\noindent and  for the  zero mode $\eta$ 
\be
{\ddot \eta}(\tau)+3h(\tau)
{\dot \eta}(\tau)+\left[1+\eta^2(\tau)+|\phi_{q_0}(\tau)|^2
\right]\eta(\tau)  =  0 \label{etacomogen}  
\ee
\noindent and the Hubble parameter given by (\ref{friedq0}) with the
energy density given by  
\be 
\epsilon(\tau) = \frac12 \, \left(|{\dot \phi}_{q_0}(\tau)|^2 +{\dot
\eta}^2(\tau)\right)  
+ \frac12 \, \left(|\phi_{q_0}(\tau)|^2+\eta^2(\tau)\right) + \frac14
\; \left(|\phi_{q_0}(\tau)|^2 +\eta^2(\tau)\right)^2 
+ \frac{q_0^2}{2 \, a^2(\tau)} \; |\phi_{q_0}(\tau)|^2 .\label{eypq0gen}
\ee
The acceleration of the scale factor is now given by
\begin{equation}
\frac{\ddot a(\tau)}{a(\tau)} = - L^2 \left[|{\dot
\phi}_{q_0}(\tau)|^2+{\dot \eta}^2(\tau) - \frac{1}{2}\left( 
|\phi_{q_0}(\tau)|^2+\eta^2(\tau)\right)-
\frac{1}{4}\left(|\phi_{q_0}(\tau)|^2+\eta^2(\tau)\right)^2  \right]
\label{accgen} 
\end{equation}

\noindent where again we have neglected the renormalization contributions and terms of ${\cal O}(g)$
 consistently in the weak coupling limit $g\ll 1$. From
eq. (\ref{accgen}) the generalized condition for an inflationary
epoch (within the  narrow tsunami case) becomes  
\be
|{\dot \phi}_{q_0}(\tau)|^2+{\dot \eta}^2(\tau) < \frac{1}{2}\left(
|\phi_{q_0}(\tau)|^2+\eta^2(\tau)\right)+
\frac{1}{4}\left(|\phi_{q_0}(\tau)|^2+\eta^2(\tau)\right)^2   
\ee
\noindent which is fulfilled if the following {\em generalized slow
roll condition} holds 
\be
|{\dot \phi}_{q_0}(\tau)|^2+{\dot \eta}^2(\tau) \ll 
|\phi_{q_0}(\tau)|^2+\eta^2(\tau) \label{slowrollgen}
\ee
Under these conditions and from eqs.(\ref{modoq0gen})-(\ref{etacomogen})
and the dynamics of the scale factor driven by the 
energy density eq.(\ref{eypq0gen})  we can now distinguish the following
different inflationary scenarios. 

\begin{itemize}

\item{{\em Tsunami dominated:} When $ (q^2_0 +1) \;
|\phi_{q_0}(0)|^2 \gtrsim \eta^2(0)$ the excited states 
in the tsunami-wave carry most of the initial energy density. In this
case the results of the previous section apply and the 
scale factor takes the form as in eq.(\ref{friango})  with $D$ given
by eq.(\ref{ctestaua}), and $E$
given by eq.(\ref{ctestaua}) but with $g\Sigma_0 \rightarrow
g\Sigma_0+\eta^2(0)$. There are {\em two} consecutive inflationary
stages as in the previous section. The first described by
eq.(\ref{soltaua}), lasts  up to the time scale $\tau_A$ defined by
$$
{\dot \eta}^2(\tau_A) + |{\dot \phi}_{q_0}(\tau_A)|^2 \sim
\frac{q_0^2}{3 \, a^2(\tau_A)} \; |\phi_{q_0}(\tau_A)|^2  
$$
at which the red-shift of the momentum $q_0$ is such that $q_0/a(\tau_A)\ll
1$. The  secondary stage is a  usual classical chaotic inflationary 
epoch determined by the dynamics of an effective zero mode given by
\be
\eta^2_{eff}(\tau) = \eta^2(\tau)+ |\phi_{q_0}(\tau)|^2 \label{etaeffgen}
\ee
\noindent because for $\tau > \tau_A$, $q^2_0/a^2(\tau)\ll 1 $ and the
effective equation of motion for $\phi_{q_0}(\tau)$ is the same as
that for $\eta(\tau)$. } 

\item{{\em Zero mode dominated:} When $\eta^2(0) \gg (q^2_0 +1) \; 
|\phi_{q_0}(0)|^2 $ the energy density stored in the
zero mode is much larger than that contributed by the excited states in the
tsunami-wave. In this case the energy density  eq.(\ref{eypq0gen}) is 
completely dominated by the zero mode. The ensuing dynamics is the
familiar  classical chaotic scenario driven by a classical zero mode, {\em
without} an early stage in which the scale factor is given by
eq.(\ref{soltaua}) which is the hallmark of the tsunami-wave dynamics.} 

\end{itemize} 

This analysis in the narrow tsunami case does highlight the
important aspects of the dynamics in a clear manner, allowing 
a clean separation of the two cases described above. We have carried a
full numerical integration of the equations of motion that 
reproduce the results described above. The criterion for the crossover
between tsunami-wave and classical chaotic inflation is determined by
the relative contributions to the energy density from  the quantum
fluctuations in the tsunami wave state as compared to the energy
density of the zero mode .   

The previous results [eqs.(\ref{dimlessexp})-(\ref{etaeffgen})] can be easily
generalized for generic continuous distributions of modes and for
mixed states. One has just to make the changes indicated in
eq.(\ref{cambio}) for generic distributions and in eq.(\ref{change})
for mixed states. 

The generalized slow-roll conditions takes then the form:
$$
{\dot \eta}^2(\tau) + g \, \int q^2 \; dq \; |{\dot f}_q(\tau)|^2 \,
 \coth\left[\frac{\Theta_q}{2}\right]\ll 
 {\eta}^2(\tau) + g \, \int q^2 \; dq \; |f_q(\tau)|^2 \,
 \coth\left[\frac{\Theta_q}{2}\right] 
$$
during the first stage of inflation. 

The dynamics is tsunami dominated provided,
$$
g \, \int q^2 \; dq \; (1 + q^2) \, |f_q(0)|^2 \,
\coth\left[\frac{\Theta_q}{2}\right]  \gtrsim \eta^2(0)
$$
The effective zero mode in the second stage of inflation is now given by
$$
{\eta}^2_{eff}(\tau) = {\eta}^2(\tau) + g \, \int q^2 \; dq \;
|f_q(\tau)|^2 \, \coth\left[\frac{\Theta_q}{2}\right]\; .
$$
These results have been verified by numerical integration of the full
set of evolution equations (\ref{pmase})-(\ref{h2tau}).

\section{Conclusions}

We have presented here inflation in typical scalar field
theories as a consequence of the time evolution of a novel quantum
state. This quantum state is characterized by a {\em vanishing}
expectation value of the scalar field, i.e,  
a vanishing zero mode, but  a non-perturbatively large number of
quanta in a momentum band, thus its name--tsunami-wave state. 

This state leads to a non-perturbatively large energy density which 
is localized in the band of excited quantum modes. 
We find that the self-consistent equations for the evolution of this 
quantum state and the scale factor lead to inflation under 
conditions that are the quantum analog of slow-roll.

The self-consistent evolution was studied analytically and numerically
in a wide range of parameters for the shape and position of the 
distribution of excited quanta. The numerical results confirm all the 
features obtained from the analytic treatment.   

Under the conditions that guarantee inflation,  there are two
consecutive but distinct 
inflationary epochs. The first stage features a rapid fall-off of the
Hubble parameter and is characterized by 
the quantum aspects of the state. During this first stage the large
number of quanta in the excited band are redshifted and build up 
an {\em effective homogeneous classical condensate}. The amplitude of
this condensate is non-perturbatively large, of ${\cal O}(1/\lambda)$,
as a consequence of the non-perturbatively large number of quanta in
the band of excited modes.  

The second stage is similar to the classical chaotic scenario and it is
driven by the dynamics 
of this effective classical condensate, with  vanishing  expectation
value of the scalar field. Under the tsunami slow-roll
conditions  on the quantum state, the total number of e-folds is more
than enough to satisfy the constraints of 
inflationary cosmology. The band of excited wave-vectors if not
initially outside the causal horizon, becomes superhorizon 
during the first inflationary stage, therefore these excited states do
not modify the power spectrum of scalar 
density perturbations  on wavelengths that are of cosmological
relevance today.   

Therefore, these tsunami-wave quantum states provide a quantum field
theoretical justification of chaotic (or in 
general large field) inflationary models and yield to a
microscopic understanding of the emergence of classical 
homogeneous field configurations of large amplitude as an effective
collective mode built from the large number 
of quanta in the excited band.

In addition, we recall that it is necessary to choose an initial state
that breaks the $ \Phi \to - \Phi $ symmetry in classical chaotic
scenarios \cite{revius,coles,linde}. This is {\em not} the case here. We
have inflation with {\em zero} expectation value of the scalar field.

For completeness we have also studied more general states and
established the important difference between tsunami (pure or mixed)
quantum states leading to inflation, and thermal mixed states which do
not lead to inflation.   

\acknowledgements
It is a pleasure to thank Daniel Boyanovsky and Francisaco Cao with
whom I have the pleasure to collaborate in this subject.

\appendix

\section{Equations of motion in the large $ N $ limit and initial states.}

 In this appendix we obtain the equations of motion in conformal time
for the generalized case in which 
the initial state is determined by a density matrix. The evolution of
the functional density matrix is given by the Liouville 
equation in conformal time
\begin{eqnarray} \label{Liouveq}
i\frac{\partial\rho}{\partial{\cal T}} &=& [H,\,\rho]\quad\Longrightarrow\cr\cr
i\frac{\partial\;}{\partial {\cal T}} \,
\rho[\vec\Psi,\,\vec{\tilde{\Psi}};\,{\cal T}] &=& 
\left( H \left[ \frac{\partial\;}{\partial\vec\Psi};\,\vec\Psi \right]
- H \left[ \frac{\partial\;}{\partial\vec{\tilde{\Psi}}};
\,\vec{\tilde{\Psi}} \right] \right)\; 
\rho[\vec\Psi,\,\vec{\tilde{\Psi}};\,{\cal T}]
\end{eqnarray}

\noindent where the Hamiltonian $H$ is given by eq.(\ref{hamk}) to
leading order in the large $N$ limit. Consistently with 
the fact that in the large $N$ limit the Hamiltonian describes a
collection of harmonic oscillators, we propose 
a  Gaussian density matrix  
\begin{equation} \label{defrho}
\rho[\vec\pi, \vec{\tilde{\pi}}, {\cal T}] = {\cal N}({\cal T})
  \prod_{k}\exp\left\{-\frac{A_{k}({\cal T})}{2}\;
    \vec\pi_{{k}}\cdot\vec\pi_{-{k}}
  -\frac{A_{k}^*({\cal T})}{2}\;
    \vec{\tilde{\pi}}_{{k}}\cdot\vec{\tilde{\pi}}_{-{k}}
  -B_{k}({\cal T})\;\vec\pi_{{k}}\cdot
  \vec{\tilde{\pi}}_{-{k}} \right\}
\end{equation}
The hermiticity condition $ \rho^\dag = \rho $ for the density matrix 
impose that $ B_{k} $ must be real.
In addition, since $ \vec\pi({\bf x}, {\cal T}) $ is a real field, its Fourier components must
obey the hermiticity condition  
$ \vec\pi_{-{k}}({\cal T})= \vec\pi^*_{{k}}({\cal T}) $; 
thus, we can assume $ A_{-{k}}({\cal T}) = A_{k}({\cal T}) $
without loss of generality.

The evolution equations for $ A_{k}({\cal T}) $, $ {\cal N}({\cal T}) $
and $ B_k({\cal T}) $ are obtained from the Liouville eq. (\ref{Liouveq})
where the hamiltonian is given by eq. (\ref{hamk}). We find
\begin{eqnarray}
i A^{'}_{k} &=& A^2_{k} -  B^2_{k} 
  - {a^2({\cal T})}\,  \omega^2_k({\cal T}) \qquad , \qquad
i B^{'}_{k} = B_{k}\,(A_{k} - A^*_{k}) 
 \cr\cr
{\cal N}_{\rho}({\cal T}) & = & {\cal N}_{\rho}(0) \; e^{-{i N \over
2}\int_0^{\cal T} d{\tilde{ \cal T}} \sum_{k}
\left[A_{k}({\tilde{ \cal T}}) - A^*_{k}({\tilde{ \cal T}})\right] } 
\; , \label{norma}
\end{eqnarray}
where the prime denotes derivative with respect to conformal time $ {\cal T} $.

The normalization factor for mixed states $ {\cal N}_{\rho}({\cal T})
$ is related with the normalization factor of pure states $ {\cal
N}_{\Upsilon}({\cal T}) $ by
$$
{\cal N}_{\rho}({\cal T}) = {\cal N}_{\Upsilon}({\cal T}) \; {\cal
N}_{\Upsilon}({\cal T})^* 
$$
where
$$
{\cal N}_{\Upsilon}({\cal T})= {\cal N}_{\Upsilon}(0) \;
\exp\left\{-i\int^{\cal T}_0{d{\cal T}' 
\left[  N V h_{cl}({\cal T}') - \frac{\lambda}{8 \, N}
\left( \sum_{k} \langle \vec\pi_{{k}}\cdot\vec\pi_{-{k}}
\rangle({\cal T}')\right)^2 + {N \over 2} \sum_{k}A_{k}({\cal T}') \right]}
\right\}
$$

Writing $ A_{k} $ in terms of its real and imaginary parts
$ A_{k} = A_{R,{k}} + i A_{I,{k}} $, we find that $ B_{k}/A_{R,{k}} $
is a conserved quantity. Thus, we can introduce  without loss of
generality the  variables $  {\cal A}_{R,{k}}({\cal T}), \;  {\cal
A}_{I,{\bf k}}({\cal T}) $ and $ \Theta_{k} $ defined by 
\begin{eqnarray} \label{defcalA}
A_{R,{k}}({\cal T}) &\equiv& {\cal A}_{R,{k}}({\cal T})
\,\coth\Theta_{k} \quad , \quad A_{I,{k}}({\cal T}) \equiv {\cal
A}_{I,{k}}({\cal T}) \cr\cr 
B_{k}({\cal T})&\equiv& -\;\frac{{\cal A}_{R,{k}}({\cal T})}{\sinh\Theta_{k}} 
\end{eqnarray}
where $ \Theta_{k} $ is a time independent  real function.

Introducing the complex variable
\begin{equation}
{\cal A}_{k} = {\cal A}_{R,{k}} + i {\cal A}_{I,{k}}
\end{equation}
we see that it obeys the following Ricatti equation
\begin{equation} \label{riccalA}
i {\cal A}^{'}_{k} = {\cal A}^2_{k} 
 - a^2({\cal T})\; \omega^2_k({\cal T})
\end{equation}
This equation can be linearized defining
\begin{equation} \label{defphik}
{\cal A}_{k}({\cal T}) \equiv - i\,
  \frac{\varphi^{'*}_{k}({\cal T})}{\varphi^*_{k}({\cal T})} \; .
\end{equation}
Then eq. (\ref{riccalA}) implies that the mode functions $ \varphi_{k} $
obey 
\begin{eqnarray} \label{eqevolmodconf}
&&\varphi^{''}_{k} +  \omega^2_ k({\cal T})\; \varphi_{k}
= 0 \; , \cr\cr 
&& \omega^2_{ k}({\cal T}) = k^2 + a^2({\cal T})\, \left[ {\cal M}^2({\cal
T}) - \frac{{\cal R}({\cal T})}{6} \right]   \label{omegadef}\; ,
\end{eqnarray}
where $ {\cal R}({\cal T}) $ is the Ricci scalar.

The relation (\ref{defphik}) defines the mode functions $
\varphi_{k}({\cal T}) $ up to an arbitrary multiplicative constant
that we choose such that the wronskian takes the value,
\begin{equation} \label{wrons}
\varphi_{k}\,\varphi^{'*}_{k} - \varphi^{'}_{k}\,\varphi^*_{k} = 2 i \; .
\end{equation}
For this choice of the Wronskian the definition (\ref{defphik}) becomes
\begin{equation}
{\cal A}_{k} = \frac{1}{|\varphi_{k}|^2} 
 -\frac{i}{2}\, \frac{d\;}{d{\cal T}}\,\ln|\varphi_{k}|^2  \; .
\end{equation}
The mass term in eq.(\ref{omegadef}) given by eq. (\ref{M2}) requires the
self-consistent expectation value  
\begin{eqnarray}
\frac{\langle\vec\pi^2\rangle_\rho}{N} &=& \int{\frac{d^3k}{(2\pi)^3}\,
  \langle\vec\pi_{{k}}\cdot\vec\pi_{-{k}}\rangle_\rho}
\cr\cr
\langle\vec\pi_{{k}}\cdot\vec\pi_{-{k}}\rangle_\rho 
  &=& \mbox{Tr}\rho \, 
  \vec\pi_{{k}}\cdot\vec\pi_{-{k}}=\frac{1}{2\,[A_{R,{k}}+B_{k}]} 
  = \frac{1}{2\,{\cal A}_{R,{k}}}\, 
  \coth\left(\frac{\Theta_{k}}{2}\right)
  = \frac12 \; |\varphi_{k}|^2\, 
  \coth\left(\frac{\Theta_{k}}{2}\right)
\end{eqnarray}
Thus, the evolution equations in terms of the mode functions are given 
by eq. (\ref{eqevolmodconf}) with
\begin{equation} \label{explmass}
{\cal M}^2({\cal T}) = m^2 + \xi\,{\cal R} 
  + \frac{\lambda}{2}\,\frac{\psi^2}{a^2}
  + \frac{\lambda}{4}\,\int{\frac{d^3k}{(2\pi)^3}\,
  \frac{|\varphi_{k}|^2}{a^2}\, \coth\left(\frac{\Theta_{k}}{2}\right)}
\end{equation}
The evolution equation of the mode  functions $ \varphi_{k} $ is the
same as  the Heisenberg equations of motion 
for the fields, hence we can write the Heisenberg field operators as
\begin{equation}
\vec\pi({\bf x},\,{\cal T}) 
  = \int{\frac{d^3k}{\sqrt2(2\pi)^3} \, \left[
  \vec{a}_{k} \; \varphi_{k}({\cal T}) \; e^{i\,{k}\cdot{\bf x}}
  + \vec{a}^\dag_{k}\,\varphi^*_{k}({\cal T})\,e^{-i\,{k}\cdot{\bf x}}
  \right]}
\end{equation}
Thus, the definition (\ref{defphik}) gives the relation between 
Schr\"odinger and Heisenberg pictures, since the functional density matrix
(\ref{defrho}) is in the Schr\"odinger picture.

The expectation value $\psi({\cal T})$ [see eq. (\ref{fieldsplit})] in
conformal time obeys the following equation of motion\cite{noscos}
\bea 
&&\psi^{''}({\cal T}) + a^2({\cal T})\, \left[ {\cal M}^2({\cal T}) 
  - \frac{{\cal R}({\cal T})}{6} \right]\;\psi({\cal T}) = 0 \label{evoleqexpconf}\\
&&\psi(0)=\psi_0 ~~;~~ \psi'(0)=\psi'_0 \label{iniconexpval}
\eea

Hence, the evolution equations  are given by
(\ref{eqevolmodconf}), (\ref{explmass}) and (\ref{evoleqexpconf}) with (\ref{iniconexpval}).

The initial density matrix in the Schr\"odinguer 
picture is  determined by  specifying the initial values of $ {\cal A}_{R,{ k}} $, 
$ {\cal A}_{I,{ k}} $ and $ \Theta_{ k} $. We will take $ a(0) = 1 $
and parameterize the initial value of $ {\cal A}_{{ k}} $ as follows,
\begin{equation} \label{schinicond}
{\cal A}_{R,{ k}}(0) = \Omega_{ k} \quad , \quad  
{\cal A}_{I,{ k}}(0) = \omega_k(0)\; \delta_{ k}
\end{equation}
The corresponding initial conditions for the mode functions 
 are obtained from eq.(\ref{schinicond}) using eq.(\ref{defphik}) and
 the Wronskian constraint  eq.(\ref{wrons}). These are given by  
\begin{equation} \label{inimodeconf}
\varphi_{ k}(0) = \frac{1}{\sqrt{\Omega_{k}}} \;; \quad\quad 
\varphi^{'}_{ k}(0) = - [\omega_k(0)\,\delta_{ k}
+ i \Omega_{ k}] \; \varphi_{ k}(0)
\end{equation}
Defining the number of particles in terms of the adiabatic eigenstates
of the Hamiltonian (\ref{hamk}) as in 
eq. (\ref{ocunumb}), it is straightforward to find that the  initial
occupation numbers are given by  
\begin{equation}
n_{k}(0) = \langle \hat{n}_{ k}(0) \rangle_{\rho(0)} =
\frac{\Omega_{ k}^2 + \omega^2_{k}(0)
+ \omega^2_{ k}(0)\,\delta^2_{k}}{4\,\omega_k\,\Omega_{ k}}\,
\mbox{coth}\left(\frac{\Theta_{ k}}{2}\right)-\frac12 \label{occumix}
\end{equation}     
For any mixing parameter $\Theta_k \neq 0$ the density matrix
represents a {\em mixed state} since $B_k \neq 0$, a {\em pure} 
initial state is obtained by  taking $\Theta_{k} = \infty$, in which
case $ B_k \rightarrow 0 $ and the density matrix becomes a 
product of a wave functional times its complex conjugate. 

It is convenient to pass to comoving time, this is achieved by the
rescaling of the fields 
\be
\psi({\cal T}(t)) = \phi(t) \; a(t)\quad , \quad
\varphi_{k}({\cal T}(t))= f_k(t) \; a(t) \label{zeromodecomo}
\ee
\noindent in terms of which  the equations of motion are 
\begin{eqnarray}
&&\ddot \phi(t) + 3\,H(t)\,\dot\phi(t) + {\cal M}^2(t) \phi(t) = 0 \cr\cr
&&\ddot f_{k}(t) + 3\,H(t)\,\dot f_{k}(t) + \left[ \frac{k^2}{a^2(t)}
+ {\cal M}^2 (t)    \right] f_{k}(t) = 0 \cr\cr
&&{\cal M}^2(t) = m^2 + \xi\,{\cal R}(t) + \frac{\lambda}{2}\,\phi^2(t) 
  + \frac{\lambda}{4}\,\int{\frac{d^3k}{(2\pi)^3}\,
  |f_{k}(t)|^2\, \coth\left(\frac{\Theta_{k}}{2}\right)} \label{equationsgene}
\end{eqnarray}
where the dots denote derivative with respect to the comoving time
$t$. The initial conditions for the order parameter are its initial value $
\phi(0) $, and its initial derivative $ \dot\phi(0) $. For $ a(0) = 1 $, 
the initial conditions for the fluctuations are given by $ \Theta_{k} $ and
\begin{equation} \label{inicondmodcom}
f_{k}(0) =  \frac{1}{\sqrt{\Omega_{k}}} \;; \quad \quad 
\dot f_{k}(0) =  - [\omega_k(0)\,\delta_{k} + H(0)
+ i \Omega_{k}] \; f_{k}(0) 
\end{equation}
[Those are the transformed of the initial conditions in conformal time 
eq. (\ref{inimodeconf}).] 

\end{document}